\documentclass[11pt,letterpaper]{scrartcl}
\usepackage{config/style}  

\title{Financial Stability Implications of \\
Generative AI: Taming the Animal Spirits
\thanks{The authors thank Jeffrey S. Allen, Marco Cipriani, Erik Heitfield, Sophia Kazinnik, Dan Li, Molly Mahar, and Nitish Sinha and participants at the Financial Stability Workshop at the Federal Reserve Board for valuable feedback. The views expressed in this paper are solely those of the authors and do not reflect the opinions of the Federal Reserve Bank of Richmond or the Board of Governors of the Federal Reserve System. Generative AI was used in the production of this paper. All errors are the authors'. Contact information: anne.hansen@rich.frb.org, seung.j.lee@frb.gov.}
\bigskip}

\author[a,b]{Anne Lundgaard Hansen}
\author[a]{Seung Jung Lee}

\affil[a]{\small\textit{Board of Governors of the Federal Reserve System}}
\affil[b]{\small\textit{Federal Reserve Bank of Richmond}}

\date{\large{\today}}

\begin{document}
\linespread{1.25}\selectfont
\maketitle
\small{
This paper investigates the impact of the adoption of generative AI on financial stability. We conduct laboratory-style experiments using large language models to replicate classic studies on herd behavior in trading decisions. Our results show that AI agents make more rational decisions than humans, relying predominantly on private information over market trends. Increased reliance on AI-powered trading advice could therefore potentially lead to fewer asset price bubbles arising from \emph{animal spirits} that trade by following the herd. However, exploring variations in the experimental settings reveals that AI agents can be induced to herd optimally when explicitly guided to make profit-maximizing decisions. While optimal herding improves market discipline, this behavior still carries potential implications for financial stability. In other experimental variations, we show that AI agents are not purely algorithmic, but have inherited some elements of human conditioning and bias. 
}
\clearpage

\linespread{1.5}\selectfont


\begin{quote}
    \noindent\textit{
    ...[T]here is the instability due to the characteristic of human nature that a large proportion of our positive activities depend on spontaneous optimism rather than mathematical expectations [...]. Most, probably, of our decisions [...] can only be taken as the result of animal spirits---a spontaneous urge to action rather than inaction, and not as the outcome of a weighted average of quantitative benefits multiplied by quantitative probabilities.}
\end{quote}
\begin{flushright}--- John Maynard Keynes\end{flushright}

\bigskip

\section{Introduction}
Human irrationality is a key driver of the build-up of financial vulnerabilities, contributing to asset price bubbles and banking crises. History offers numerous examples, including Tulip Mania in the 17th century, the South Sea Bubble, the dot-com boom, the 2008 financial crisis, the 2010 Flash Crash, and the GameStop short squeeze. A well-established body of research highlights the role of psychological and emotional factors, coined \emph{animal spirits} or \emph{irrational exuberance}, in these periods of boom and bust \citep{shiller2005irrational,degrauwe2012lectures,angeletos2018quantifying}.

Understanding the role of animal spirits for financial stability is already a challenge, given the unpredictable nature of human behavior. Now, a new and unknown agent has entered the equation: decisions powered by generative AI. Humans increasingly rely on AI for information gathering and decision making, whether as a co-pilot or as autonomous agents.\footnote{ \citet{hartley2024labor} conducted a survey covering the U.S. labor market showing that the percentage of workers adopting LLMs steadily increased from 30\% to 45\% between end of 2024 and mid 2025. These adoption rates are massive in the context of an earlier survey from 2018 showing that less than 6\% of firms used any AI-related technology, i.e., automated-guided vehicles, machine learning, machine vision, natural language processing, or voice recognition \citep{mcelheran2023ai}.} As generative AI is reshaping workflows across institutions and individuals, the question arises: How might the increased reliance on generative AI impact financial stability? Specifically, will generative AI exaggerate or dampen the role of animal spirits in the build-up of financial vulnerabilities?

Two competing hypotheses emerge. On the one hand, AI is fundamentally algorithmic, operating in a set of instructions and grounded in logic and rational decision making.\footnote{This algorithmic foundation suggests that AI systems may enhance human judgment by providing consistent assessments less prone to biases. The seminal paper by \citet{kleinberg2018human} argues that machine-learning algorithms can improve human decision making in the context of bail decisions, especially if carefully integrated into an economic framework. Similarly, \citet{li2024effect} demonstrates how AI-enabled credit scoring models can increase loan approval rates for under-served populations while simultaneously reducing default rates, primarily through the algorithmic processing of weak signals that humans might overlook or inconsistently evaluate.} If AI-guided decisions replace human intuition, the result could be a reduction in the influence of animal spirits, leading to more stable financial markets. On the other hand, generative AI models, such as large language models (LLMs), are trained on vast amounts of data, sourced from both rigorous materials, such as academic research, and the, at times, chaotic discourse of social media platforms such as Twitter (X) and Reddit. Consequently, generative AI may inherit and even amplify human biases and irrational tendencies \citep{koralus2023humans, hayes2024relative,zhu2024incoherent,jiang2023personallm}. Moreover, many AI models undergo reinforcement learning from human feedback (see \citealp{wang2024survey} for a survey), optimizing for engagement and persuasion rather than pure rationality. This suggests that instead of mitigating the build-up of financial vulnerabilities, AI could exacerbate financial turbulence driven by animal spirits. Finally, \citet{danielsson2024ai} argue that AI adoption will likely cause more intense future crises due to AI's ability to respond quickly to shocks. The net effect of AI's involvement in financial decision making is therefore unclear. 

This paper explores these competing perspectives, examining the implications of the expanding role of AI in economic decision making for financial stability. We focus on the potential financial vulnerabilities driven by irrational tendencies of decision makers in financial markets, and more specifically on herd behavior. Herd behavior---where investors ignore private signals and mimic the decisions of others, potentially driving prices away from fundamental values---is a well-documented form of irrationality that can cause asset price bubbles \citep{galariotis2016herd,hsieh2020retail}. Even in cases where herding is optimal, this type of behavior can contribute to financial stability events by increasing market volatility and accelerating price movements which can reverse quickly in the event of new information arriving \citep{bikhchandani2001herd,chamley2003rational}. 

We conduct laboratory-style experiments using LLMs to replicate classic studies on herd behavior in trading decisions. These experiments provide novel insights into the behavioral tendencies of AI agents, laying a micro-foundation for future work on financial stability in an AI-powered economy. Our micro-level approach is motivated by \citet{horton2023large}, who argues that LLMs can be treated as agents whose decisions and behavior can be studied in parallel to studies of human behavior. We stress that we are not attempting to model a realistic financial market, but rather to zoom in on the behavior of LLMs in a controlled setting.\footnote{While a complementary study of financial stability implications of generative AI for macro-level outcomes is certainly of interest, we conjecture that such an undertaking would be difficult to achieve. First, there are the usual problems of disentangling effects from human behavior from confounding factors and distinguishing intentional herding from situations where the private information of many traders happen to coincide, coined as ``spurious herding'' in \citet{bikhchandani2001herd}. Second, the world is still in early stages of generative AI adoption with impacts yet to be seen.} In particular, our setting allows us to observe detailed data on each decision made by AI agents, including their reasoning. 

We focus on the experiment by \citet{cipriani2009herd}, which investigates herd behavior among 32 financial market professionals through a controlled laboratory setting. Their setting is grounded in an established model of herd behavior \citep{avery1998multidimensional}, which has been rigorously tested in laboratory-style experiments \citep{cipriani2005herd,drehmann2005herding}. Whereas the previous experimental literature conducted experiments with undergraduate students, \citet{cipriani2009herd} recruited financial market professionals. These details makes their study particularly interesting as a human benchmark for our purpose. After all, it is the actions of financial market professionals---not students---that shape real-world market dynamics and impact the stability of the financial system. 

In this setting, herd behavior arises when investors disregard private information to follow market trends.\footnote{A related concept is \textit{momentum}, which describes a situation where high-return stocks continue to exhibit high returns. While herding and momentum are clearly related, they are inherently different: herding refers to behavior relative to investors' private information, whereas momentum involves behavior relative to past prices only.} This definition is narrower than those found in broader and recent discussions of generative AI adoption and herding. For example, \citet{danielsson2024ai} and a report from the \citet{fsb2024financial} raise the issue that wide-spread AI-usage among financial market participants may streamline modeling approaches leading to increased market correlation. Herding is also fundamentally different from the concept of collusion, which is studied in the context of AI adoption in \citet{dou2025ai} who conduct simulated experiments to prove that autonomous reinforcement learning algorithms collude by coordinating their trading decisions to earn supra-competitive profits. Collusion involves strategic and deliberate coordination. In contrast, herding is uncoordinated imitation arising from informational spillovers or psychological biases. While concerns of model monoculture and collusion are certainly relevant to consider, they are not the focus of our work. We aim to provide insights into behavioral aspects of LLMs and their tendency to herd in trading decisions, and leave the impact of AI adoption on the dynamics among market participants for future research.

We replicate the \citet{cipriani2009herd} experiments using trading decisions of LLMs (which we refer to as \emph{AI agents}) in place of decisions made by human participants. Our implementation closely following the original experimental design. Specifically, we prompt the LLMs with instructions that mirror those given to financial professionals in the human study. This setup allows us to compare LLM decision making with the human results taken directly from \citet{cipriani2009herd}: the \quotation{AI laboratory} versus the \quotation{human laboratory.} To generalize our results as much as possible, we use four different LLMs\footnote{These LLMs are: Anthropic's Claude 3.5 Sonnet and Claude 3.7 Sonnet models, Meta's Llama 3 Instruct 70B model, and Amazon's Nova Pro model.} and average the results across models. 

Our results show that AI agents demonstrate significantly more rational trading decisions, which are decisions based on private information, compared to human participants. Across different parameterizations of the experiment, AI agents made rational decisions between 61-97\% of the time, substantially exceeding the 46-51\% range observed in human participants. The AI laboratory also exhibited fewer information cascades, where investors trade based on the actions of other investors rather than relying on their own private information. Specifically, information cascades occurred between 0-9\% of the AI decisions, compared to around 20\% for humans. Notably, when AI agents did engage in cascade trading behavior, they traded against market trends (contrarian behavior) rather than following the herd. In addition, we show that AI agents do not exhibit completely irrational behavior (or make trading errors), which contrasts with the results of the experiments conducted with human participants. We interpret these results as early indications that a future where investors are more impacted by advice generated by LLMs can potentially involve fewer asset price bubbles arising from herd behavior. 

However, studying the rationals provided by LLMs alongside the trading decisions reveals that AI agents rely too heavily on their private information. As market trends can reflect the private information of others, it can be optimal from a profit-maximizing perspective to take trading history into account when making trading decisions. We show that AI agents fail to acknowledge this, leading to occasional suboptimal choices. We therefore implement AI agents that are prompted with additional guidance on optimal decision making; instructions that their human counterparts did not receive in the original field experiment. This experiment serves as a proxy for the fine-tuning that financial institutions would likely implement if adopting generative AI for powering trading. We show that these optimal AI agents do engage in cascade trading when optimal, but that they still remain reluctant to herding.

Finally, we explore variations of the experiment to examine whether different conditions lead to stronger evidence of herd behavior. Unlike the original study, scaling up the length of the experiments and modifying their parameters is both cost-effective and efficient with LLMs. For example, we test the impact of re-labeling the private signals that participants receive during the experiment. The original experiment uses neutral color-coding for these signals. We test outcomes using non-neutral colors, such as green and red. Using \quotation{green} to indicate a high probability of a high asset value and \quotation{red} to indicate a high probability of a low asset value yield similar results as the baseline experiment. However, reverting the labeling such that \quotation{red} (\quotation{green}) signals a high probability of a high (low) asset value, which is counterintuitive given human conditioning, the LLMs generate very few rational responses. AI agents are therefore not algorithmic rational, following a well-defined set of rules, but has inherited some elements of human intuition and bias. This finding is consistent with a growing literature showing that LLMs can replicate human errors and biases \citep{argyle2023out,bybee2023ghost,hansen2025simulating}.




Our results suggest that AI agents exhibit less herd behavior than human financial professionals, a finding with significant implications for future financial stability as generative AI gains traction in market decision making. Specifically, the reduced tendency to herd could potentially lead to less extreme market movements and fewer asset price bubbles, contributing to greater overall financial market stability. However, we acknowledge that the introduction of AI agents could fundamentally alter market dynamics in ways that are not yet fully understood. Continued research and adaptive regulatory approaches to maintain financial stability in an AI-augmented financial landscape is therefore warranted.

We proceed as follows. Section \ref{sec:literature} reviews the literature. Section \ref{sec:herding_financialstability} discusses the concept of herding and its relation to financial stability. Section \ref{sec:experimental_design} outlines the theoretical model that underpins the experimental design. Section \ref{sec:laboratory_setup} describes the human laboratory in which the experiment was conducted in \citet{cipriani2009herd}, and how we adopt this setting with LLMs. Section \ref{sec:results} presents the main results. In Section \ref{sec:extensions}, we introduce various alterations to the experiment to understand the prevalence of herd behavior under different experimental settings. Section \ref{sec:discussion} provides an overview and discussion of our main results. Conclusions follow in Section \ref{sec:conclusion}.

\section{Literature}\label{sec:literature}
This work contributes to the growing literature on the behavior of LLMs. While our study focuses on herd behavior and financial stability, other works have examined other types of behavior and departures from rationality. Most relevant to our approach is \citet{henning2025llm}, who conducts asset pricing experiments with LLM traders demonstrating that AI agents tend to price assets near fundamental values. As in our work, they conclude that AI adoption has the potential to enhance financial stability by dampening the likelihood of asset price bubbles. Their test is, however, fundamentally different from ours. In \citet{henning2025llm}, agents choose whether to invest cash in a risky asset under full information about expected dividends and interest rates, which directly facilitates computation of the fundamental value. This setup seeks to test rationality in the context of price setting rather than behavioral aspects of asset price bubbles, which is our focus. 

\citet{chen2023emergence} studies the economic rationality of GPT models by conducting revealed preference experiments, where models are prompted to make decisions under budget constraints. Similar to our results, although in a different aspect of the term rationality, the authors conclude that AI agents tend to exhibit more rational behavior than humans. \citet{liu2025large} confirms that LLMs are are more rational than humans using a large data set of human decisions in risky choice problems. \citet{riochanona2025can} focuses on laboratory experiments related to price expectations and deviations from rational expectations. They emphasize the importance of the interactions of different AI agents and retaining memory across time periods; both elements that we include into our AI laboratory setting as well. While they conclude that LLMs are not strictly rational in their expectation formation, they find that LLMs generate less variability in their responses compared with humans. Similar patterns are observed in our results.  
 
While these studies, like ours, mainly emphasize differences between the behaviors of humans and AI agents, some papers emphasize their similarities and argue that LLMs can be used to simulate human outcomes. For example, \citet{horton2023large} argues that LLMs can give human-like responses and suggests that they can be used conduct pilot experiments to calibrate experimental designs before testing on human beings. \citet{hansen2025simulating,jha2024chatgpt}, and \citet{zarifhonarvar2024experimental} show that LLMs can be used to simulate economic surveys. The literature emphasizes that LLMs in some contexts exhibit human biases such as risk aversion and loss aversion \citep{jia2024decision,ross2024llm}. Along the same lines, \citet{hua2024game} show that LLMs often deviate from rational decisions in game theoretic experiments. Characterizing the exact distinction between human and AI decision making remains an open question.

Finally, our work is particularly important as LLMs start to play an increasingly larger role in financial market decisions. The literature showcases the application of LLMs in investing. \citet{lopezlira2025can} simulates a stock market using LLMs and argues that this framework can be used to conduct counterfactual experiments. \citet{lee2025your} argues that LLMs suffer from confirmation bias in the realm of investment. Specifically, the authors show that LLMs exhibit a preference for large-cap stocks and contrarian trading strategies. And \citet{fedyk2024ai} surveys the investment preferences of human and AI agents, finding that AI demonstrates demographic biases that can be overcome with demographically-seeded prompts.

\section{Herding and financial stability}\label{sec:herding_financialstability}
The literature distinguishes between optimal and suboptimal herding, each with distinct implications. Optimal herding represents a fundamental mechanism through which individually optimal decisions can generate collective fragility in financial markets. In the canonical models of \citet{bikhchandani1992theory} and \citet{banerjee1992simple}, optimal herding occurs when mimicking others represents an optimal response to superior information possessed by early movers. Once sufficient agents have acted in one direction, subsequent investors optimally ignore their own private signals, leading to collective behavior that may diverge from fundamental values. 

Suboptimal herding occurs when investors follow the crowd even when this leads to lower expected profits than decisions relying on private information. This behavior stems from cognitive biases, reputational concerns \citep{scharfstein1990herd}, and misaligned incentives. One recent example of suboptimal herding behavior is the "meme stock mania" of 2021, exemplified by GameStop and AMC, where retail investors collectively moved markets by following social trends rather than fundamental valuations.

As depicted Figure \ref{fig:herding_diagram}, both types of herding can lead to financial instability, though through different mechanisms. Optimal herding can accelerate information aggregation and enhance efficiency by incorporating dispersed knowledge into prices \citep{chamley2003rational}. For example, withdrawals from a genuinely insolvent bank reflect individually rational and collectively justified behavior. Such market discipline may improve financial stability if the market is able to distinguish between insolvent banks and solvent banks. Nevertheless, even optimal herding may generate short-term volatility by accelerating price corrections, which can reverse quickly if new information arrives. Short-term volatility can also arise from concentrating trading flows, potentially exacerbating fire sales and liquidity strains. Finally, optimal herding can overshoot or undershoot fundamental values, or transmit to other markets, triggering suboptimal herding with adverse consequences \citep{cipriani2008herd}. These dynamics create a paradox: actions that optimize individual utility can simultaneously undermine market efficiency and stability. Financial stability may therefore be enhanced when investors act on their private signals (which we refer to as rational behavior) rather than engage in behavior that is individually optimal but systemically destabilizing.

Suboptimal herding proves particularly destabilizing as it transforms noise into crises—-prices severely deviate from fundamentals, liquidity evaporates, and self-fulfilling runs emerge. Suboptimal herding can even lead to contagion. The 2008 freezing of interbank and other markets exemplifies this dynamic, where even solvent counterparties lost funding access amid generalized panic. By suppressing diverse private signals, suboptimal herding fundamentally undermines the information efficiency of markets.


For financial stability, the distinction between herding types carries significant policy implications. Suboptimal herding can be mitigated through enhanced transparency, reformed incentives, and market infrastructure improvements (e.g., circuit breakers), as it originates in behavioral amplification rather than fundamental weaknesses. Optimal herding, however, proves more challenging to address without resolving underlying vulnerabilities such as undercapitalization or asset toxicity. Both variants contribute to financial fragility through fire-sale externalities, liquidity spirals, and cross-institutional contagion \citep{brunnermeier2009market}. While herding can occasionally enhance efficiency through information aggregation, its tendency to suppress private signals renders it a persistent source of systemic risk. 

\section{Experimental design}\label{sec:experimental_design}
This section presents the model and theoretical predictions outlined in \citet{cipriani2009herd}. The model is based on \citet{avery1998multidimensional}.

\subsection{Theoretical model}
The model describes a financial market with one risky asset and discrete trading periods indexed by $t=\{1,2,\hdots\}$. Before the first trading period, there is a $\rho$ probability of an \textit{information event}, which changes the fundamental value of the asset in either direction. In each trading period, some traders are informed and receive a private signal on the value change, while others do not. The model characterizes different types of trading behaviors based on whether informed traders act according to their private signal (\textit{rational} behavior) or ignore their signal (\textit{cascade} behavior).

The asset's fundamental value belongs to the discrete set $v\in\curly{0,50,100}$. Specifically, if there is no information event (with probability $1-\rho$), the value is equal to its unconditional expected value given equal probabilities, i.e., $v=50$. An information event (occurring with probability $\rho$) pushes the value to zero or 100, with the following probability distribution: $\Pr{v=0}=\Pr{v=100}=0.5$. The asset trades at a price $p$, which is set by the market maker according to Bayesian updating as we detail below.

Traders act sequentially with only one trader randomly chosen to trade in each trading period.\footnote{This structure simulates the mechanics of central limit order book trading.} In each period $t$, the chosen trader executes an action $x_t$, which is to buy one unit of the asset ($x_t=\text{buy}$), sell one unit of the asset ($x_t=\text{sell}$), or not trade ($x_t=\text{no trade}$). If there is no information event, all traders are uninformed \emph{noise traders}, who trade based on exogenous probabilities, i.e., $\Pr{x_t=\text{sell}}=\Pr{x_t=\text{buy}}=\Pr{x_t=\text{no trade}}=1/3$. In the case of an information event, the chosen trader is \emph{informed} with probability $\mu$ (and a noise trader with probability $1-\mu$). An informed trader receives a signal $s_t\in\curly{\text{white},\text{blue}}$, which is tied to the asset value in the following way:
\begin{align}
    \Pr{s_t=\text{white}\mid v=100} = \Pr{s_t=\text{blue}\mid v=0} = 0.7.  
\end{align}
That is, a white signal can be interpreted as a \emph{good} signal, indicating that the information event resulted in a high asset value, whereas a blue signal is \emph{bad} in the sense that it increases the probability of a zero asset value. In addition to the signal $s_t$, an informed trader also observes the trading history $h_t$, and therefore forms beliefs about the asset value based on the conditional expected value given $s_t$ and $h_t$: $\E(v\vert s_t,h_t)$. The realized payoff is equal to $v-p$ if the trader chooses to buy the asset, $p-v$ if the trader chooses to sell, and zero if the trader chooses not to trade. We assume that the informed trader is risk-neutral and seeks to maximize expected payoff given $s_t$ and $h_t$.

A market maker facilitates exchanges with the traders and sets the price of the asset given the history of trades for periods up to $t-1$, $h_t=\{x_1,x_2,\hdots,x_{t-1}\}$ for $t>1$. $h_1=\emptyset$. Specifically, the price is determined as the expected asset value given $h_t$: $p_t=\E\paran{v\mid h_t}$.\footnote{There is only one asset price, i.e., the model assumes a zero bid-ask spread. This assumption was imposed by \citet{cipriani2009herd} to simplify the laboratory experiment.} In the first trading period, with no trading history, the price is equal to its unconditional expected value: $p_1=\half100=50$. At $t>1$, the price is given by the expected asset value conditional on the history of trades: 
\begin{align}
p_t &= 100\, \Pr{v=100\vert h_t} + 0\, \Pr{v=0\vert h_t} = 100q_t
\end{align}
where $q_t = \Pr{v=100\vert h_t}$ is determined using Bayesian updating:\footnote{The term $(1-\rho\mu)\frac{1}{3}$ represents the probability a buy or sell comes from a noise trader, who buys, sells, and chooses not to trade with equal probability. The term $\mu\rho$ is the probability that a trader is informed, given by the probability that an information event occurred ($\rho$) times the probability that a trader is informed given an informed event ($\mu$).}
\begin{align}
    q_t = &\Pr{v=100\vert x_{t-1},h_{t-1}} \\
    \begin{split}= 
    &\1{(x_{t-1}=\text{buy})} \bracket{\frac{\paran{0.7\rho\mu+(1-\rho\mu)\frac{1}{3}}q_{t-1}}{\paran{0.7\rho\mu+(1-\rho\mu)\frac{1}{3}}q_{t-1}+\paran{0.3\rho\mu+(1-\rho\mu)\frac{1}{3}}\paran{1-q_{t-1}}}} + \\
    &\1{(x_{t-1}=\text{sell})} \bracket{\frac{\paran{0.3\rho\mu+(1-\rho\mu)\frac{1}{3}}q_{t-1}}{\paran{0.3\rho\mu+(1-\rho\mu)\frac{1}{3}}q_{t-1}+\paran{0.7\rho\mu+(1-\rho\mu)\frac{1}{3}}\paran{1-q_{t-1}}}} + \\
    &\1{(x_{t-1}=\text{no trade})}q_{t-1}.\label{eq:qt}
\end{split}\end{align}    

\subsection{Theoretical predictions}
This section presents the theoretical predictions for how informed traders act according to the model. Informed traders make decisions by comparing the price of the asset to the expected value given the signal and trading history:
\begin{align}
    x_t = \begin{cases}
        \text{buy} & \text{if}\quad p_t < \E\paran{v\vert s_t,h_t}\\
        \text{sell} & \text{if}\quad p_t > \E\paran{v\vert s_t,h_t}\\
        \text{indifferent} & \text{if}\quad p_t = \E\paran{v\vert s_t,h_t}
    \end{cases}.
\end{align}
When indifferent, traders may buy, sell, or not trade; their payoff will be the same regardless of their action. Their expected value is given for each signal as follows:
\begin{align}
    \E\paran{v\vert s_t=\text{white},h_t} &= 100\bracket{\frac{0.7q_t^\star}{0.7q_t^\star+0.3(1-q_t^\star)}},\\
    \E\paran{v\vert s_t=\text{blue},h_t} &= 100\bracket{ \frac{0.3q_t^\star}{0.3q_t^\star+0.7(1-q_t^\star)}} 
\end{align}
where $q_t^\star=\Pr{v=100\vert x_{t-1},h_{t-1},\rho=1}$, which can be computed from \eqref{eq:qt} above. For the informed trader, the relevant probability of the high asset value conditional on the trading history sets $\rho=1$ because the informed trader, by definition, knows with certainty that an information event occurred. The discrepancy between $q_t$, the probability of a high asset value from the perspective of the market maker, and $q_t^\star$, the corresponding probability from the perspective of an informed trader, can lead to optimal information cascades.  

The model characterizes different types of behavior of informed traders, defined as follows:

\begin{description}
    \item[Rational:] The informed trader chooses to buy upon receiving a white (\textit{good}) signal and sell upon received a blue (\textit{bad}) signal.
    \item[Partial rational:] The informed trader follows rational behavior upon receiving one signal and to not trade upon receiving the other signal, e.g., buy upon receiving a white (\textit{good}) signal and no trade upon received a blue (\textit{bad}) signal.
    \item[Cascade trading:] The informed trader chooses the same trading action (buy or sell) regardless of the private signal. If the trader chooses to buy (sell) when the trading history is dominated by buy-actions (sell-actions), i.e., act following the majority action of previous traders, the trader engages in \textit{herd behavior}. If the trader chooses to buy (sell) when the trading history is dominated by sell-actions (buy-actions), i.e., acting against the majority of previous traders, the trader engages in \textit{contrarian behavior}.
    \item[Cascade no trading:] The informed trader chooses not to trade regardless of the private signal.
    \item[Error:] The informed trader chooses to buy upon receiving a blue (\textit{bad}) signal and sell upon receiving a white (\textit{good}) signal.
\end{description}
The last type of behavior is always sub-optimal and is interpreted as an error if observed. However, it can be optimal for traders to engage in cascade behavior, depending on the parameterizations of the model. The laboratory experiments in \citet{cipriani2009herd} follow two different parameterizations of the model, referred to as treatments. 

In the first treatment (Treatment I), there is no uncertainty about whether an information event occurs, i.e., $\rho=1$. In addition, all traders are informed, i.e., $\mu=1$. Hence, $q_t=q_t^\star$, and it follows that:
\begin{align*}
\E(v | s_t = \text{white},h_t) = 100 \,\frac{0.7q_t}{0.7q_t + 0.3(1-q_t)} > 100q_t = p_t
\end{align*}
and
\begin{align*}
\E(v | s_t = \text{blue},h_t) = 100 \, \frac{0.3q_t}{0.3q_t + 0.7(1-q_t)} < 100q_t = p_t.
\end{align*}
Hence, regardless of the history of trades, a trader's expected value given their private signal is always on the same side of the market price as their signal. Therefore, it is always optimal for traders to follow their private signals. As a result, each trade reveals new information, continuously updating the market price. This prevents the formation of information cascades, as traders never have an incentive to ignore their private information in favor of following the actions of others.

In the second treatment (Treatment II), there is uncertainty both about whether an information event occurs and the proportion of informed traders. \citet{cipriani2009herd} set $\rho=0.15$ and $\mu=0.95$, i.e., an information event occurs with 15\% probability and the probability that a trader receives a private signal on the information event is a slightly smaller than one. 

With event uncertainty, it can be optimal for traders to engage in cascade behavior. The reason is that there is information asymmetry between informed traders and the market maker. Upon receiving a private signal, the informed trader knows with certainty that an information event has occurred and that the history of trades comes from an informed trader with probability $\mu=0.95$. In contrast, not knowing whether an information event has occurred, the market maker believes that the traders are informed with probability $\rho\mu=0.15\cdot0.95=0.14$. This asymmetry leads the market maker to update the asset price more conservatively than informed traders update their beliefs. After a sequence of buy orders, the gap between traders' expectations and the market price can widen. Eventually, a trader's expectation may exceed the market price even with a contradictory signal: $\E(v|s_t=\text{white},h_t) > \E(v|s_t=\text{blue},h_t) > p_t$. At this point, the trader will ignore their private information and follow the herd.\footnote{Optimal herding behavior is temporary. When traders herd, the private signals are not reflected in the prices. However, the market maker continues to update beliefs about whether an information event has occurred, causing prices to keep moving, albeit slowly. Eventually, the price may move enough to make private information relevant again, breaking the herd behavior.} However, because the market maker updates his expectation by less than the informed traders, it will never be the case that, after a history of buys, the expectation of a trader will be below the price for both signal realizations, i.e., $p_t > \E(v|s_t=\text{white},h_t) > \E(v|s_t=\text{blue},h_t)$. As a result, an informed trader will never engage in contrarian behavior. Analogous arguments apply to a sequence of sell orders.

At the extreme, the market maker does not update the price at all such that the price remains at the unconditional expected value throughout all trading periods. \citet{cipriani2005herd} conducted an experiment with this setting (without event uncertainty) among undergraduate students. We shall refer to this setting as Treatment III. In this parametrization, optimal herding arises when there is a trade imbalance greater than or equal to two 
\citep{bikhchandani1992theory}; see \citet{cipriani2005herd} for intuition. Since this experiment was not conducted among financial market professionals, we shall focus less on this parametrization in our results.


\paragraph{Optimal behavior:} To summarize, the model predicts the following behavior in the two treatments:
\begin{description}
    \item[Treatment I (price updating; no event uncertainty):] Traders always trade according to their private signal, preventing the formation of cascades.  
    \item[Treatment II (price updating; event uncertainty):] An information cascade occurs with positive probability. Herding is optimal when prices are below the expected value conditional on both signals, but it is never optimal to engage in contrarian behavior.  
    \item[Treatment III (no price updating; no event uncertainty):] Herding is optimal after a trade imbalance higher than or equal to two.
\end{description}

\section{Laboratory setup}\label{sec:laboratory_setup}
\citet{cipriani2009herd} implemented the experiment among financial market professionals. We adopt their \emph{human laboratory} setting as closely as possible, replacing human participants with AI agents. Then we compare our results from this \emph{AI laboratory} with the human results from \citet{cipriani2009herd}. This section describes the human and AI laboratories.

\subsection{Human laboratory}
The human experiment was conducted with 32 participants working for financial institutions in London. The participants were divided into four groups of eight; each group formed one session. 

In each of the four sessions, the experiment was repeated for two practice rounds followed by first eight rounds implemented with the parametrization in Treatment I and then eight rounds with the Treatment II parametrization. Before each treatment, participants were given written instructions. They were informed that everyone received the same set of instructions, and were given the opportunity to ask clarifying questions which were answered privately. The timeline for each session was as follows:

\paragraph{Timeline for each session in human laboratory:}
\begin{enumerate}
    \item Participants were given written instructions for Treatment I. 
    \item Practice round consisting of two trading periods with Treatment I parametrization.
    \item Treatment I round consisting of eight trading periods.
    \item Participants were given written instructions for Treatment II. 
    \item Treatment II round consisting of eight trading periods.
    \item Payoffs were paid out.
    \item Participants filled out a survey collecting personal characteristics (gender, age, education, work position, job tenure). \citet{cipriani2009herd} report the unconditional distributions of these characteristics.
\end{enumerate}

Each round proceeded as follows:

\paragraph{Timeline for each round in human laboratory:}
\begin{enumerate}
    \item A computer selected the asset's fundamental value from the distribution $\Pr{v=0}=\Pr{v=100}=0.5$. In Treatment II, there is a theoretical 85\% probability that an information did not occur, leaving the value at $50$. However, the experiment was implemented \textit{as if} an event did occur. 
    \item Not knowing the asset's value $v$, participants chose their actions conditional on observing a white and blue signal.
    \item A computer randomly chose one trader from a uniform distribution, who was selected to trade. The computer also chose the realized signal from the signal's probability distribution conditional on the value selected in step 1. 
    \item The selected trader received the realized signal. The remaining traders only observed the executed action (buy, sell, no trade).
    \item The price for the next round was computed given the selected trader's action for the realized signal.
    \item Steps 2-6 were repeated for eight rounds total, until all participants had been selected to trade exactly once. 
    \item Payoffs for the round were revealed to each participant. Participants who bought (sold) the asset in the round at the price $p_t$ received $v-p_t$ ($p_t-v$) lire, a fictional currency that was translated into GBP at the end of the experiment at the exchange rate of three lire per GBP.  
\end{enumerate}

We refer to \citet{cipriani2009herd} for further details.

\subsection{AI laboratories}\label{sec:ai_laboratory}
We adopt the human experiment in our AI laboratory, where human participants are replaced by AI agents. To model AI agents, we use a suite of LLMs and apply model averaging to get an all-compassing view of the behavioral patterns of AI-powered trading. Specifically, we use Anthropic's Claude 3.5 Sonnet and Claude 3.7 Sonnet models (hereafter Claude 3.5 and 3.7), Meta's Llama 3 Instruct 70B parameter model, and Amazon's Nova Pro model. We mainly implement the models with a moderate temperature of 0.7, balancing creativity with determinism.\footnote{The temperature adjusts how the model weighs its prediction for the next token. A lower temperature makes the model focus more on its top choices, while a higher temperature gives it more freedom to consider less likely options, affecting how predictable or creative the output becomes.} Robustness checks confirm that the choice of temperature does not impact the conclusions of our experiments. The Claude 3.7 model is implemented in \emph{extended thinking mode}, requiring a temperature of 1.0. This setting activates extended reasoning capability, where the model iterates in multiple steps to reach the \quotation{best} perceived answer. 

We follow the setup of the human laboratory described above as closely as possible. For example, similar to human participants, we presented an LLM (Claude 3.5) with written instructions and gave the model the opportunity to ask clarifying questions. We used this model feedback to improve the instructions. However, some adjustments are necessary to accommodate differences between human and AI agents. First, practice rounds are redundant, and we completely separate the two treatments to avoid confusing the models. Second, we explicitly provide memory to the AI agents in each trading period, by listing the executed trades along with the history of actions and reasoning for each agent in all previous periods.

LLMs are instructed through prompts. The \emph{user prompt} sets the task or query that the user wants the model to respond to, and it can change with each interaction. In addition to the user prompt, LLMs can also be instructed through their \emph{system prompt}, which sets the context, behavior, knowledge base, and role for the model. We use the system prompt to provide the general instructions of the experiment (corresponding to the written instructions handed out to human participants) and the user prompt to provide updates throughout the trading periods and request trading actions. 

\paragraph{Timeline for each round in the AI laboratory:}
\begin{enumerate}
    \item A computer selected the asset's fundamental value, as in the human experiment.
    \item We make an API call to an LLM, using the instructions of the experiment as the system prompt, see Prompt \ref{prmpt:system}. The user prompt requests the model to provide a trading action (buy, sell, no trade) given each signal (blue and white) and the current asset price, along with its reasoning for each action. For trading rounds $t>1$, the user prompt also provides, for each agent, the history of executed trades, a notification if that agent was chosen to act in the previous round, and the history of actions and reasoning of that agent. The user prompt is provided in Prompt \ref{prmpt:user}.
    \item The trader selected trade and the realized signal are chosen by a computer, as in the human experiment.  
    \item The price for the next round is computed given the selected trader's action for the realized signal.
    \item Steps 2-4 are repeated for eight trading periods total.  
\end{enumerate}
This timeline is summarized in flow diagrams in Figure \ref{fig:flow_diagrams} for each treatment.

Each experiment (i.e., four sessions of eight trading rounds) is repeated across different LLMs. To maintain comparability across experiments, we seed the randomness such that the realized asset value, realized signals, and the sequence of selected traders are identical across the experiments. Following \citet{cipriani2009herd}, we assume that an information event always happens, even in Treatment II, where the theoretical probability of an information event is less than one. 

\paragraph{Optimal AI laboratory}
To establish a baseline, our main AI laboratory does not instruct the LLMs about what constitutes optimal decision making. In contrast, real-world adoptions of AI in financial market decision making likely attempts to guide the models in optimal behavior to the largest extent possible. We therefore also explore an optimal version of the AI laboratory, where we explicitly tell the models when herding is optimal according to theory through the user prompt. This prompt is given in Prompt \ref{prmpt:user_optimal}.

\section{Results} \label{sec:results}
This section presents the results from conducting the \citet{cipriani2009herd} experiments in the AI laboratories. We mainly focus on the baseline AI laboratory in which we do not include any guidance on optimal decision making in the prompt. AI agent decisions are presented alongside human decisions from the original experiment. First, we consider the case without event uncertainty (Treatment I), then we introduce event uncertainty (Treatment II), and finally we prevent price updating (Treatment III). We also analyze the descriptions of reasoning provided by the LLMs to understand the decisions of AI agents. Unfortunately, we do not have human counterparts for these insights. 

These baseline results establish patterns of decision making in general-purpose technology without further optimization. In real-world applications, professional traders would likely consult tools that have been fine-tuned to maximize profits according to some risk-profile. To examine herding behavior in such optimized AI agents, we consider an AI laboratory where the LLMs are informed about optimal decisions through the prompt. Finally, we consider robustness to the model temperature parameter.    

\subsection{Without event uncertainty}
We begin by discussing results obtained with the parameterization in Treatment I, where there is no model uncertainty. The theoretical model predicts that traders should always trade according to their private signals, which precludes the formation of cascades.

Table \ref{tab:table1_baseline} shows the frequency of the different behaviors averaged across all sessions and trading periods. The \quotation{Human} column recites the results from the human laboratory reported in \citet{cipriani2009herd}. The \quotation{AI} column represents the average results across all considered LLMs. With Treatment I, reported in panel (a), AI agents exhibit more rational behavior, i.e., buy on a \quotation{good} signal and sell on a \quotation{bad signal,} (61\%) compared to humans (46\%). This result is largely driven by the Claude 3.7 and Llama 3 models, which generate rational responses for a vast majority of sessions and trading periods. In contrast, the Claude 3.5 and Nova Pro models have fewer rational responses, but a majority of responses that are partially rational, i.e., follow the rational response on one signal but decide to not trade on the other signal. As a result, the share of rational and partial rational responses in the AI laboratory far exceed that observed in the human laboratory (90\% for AI versus 65\% for humans). It is worth noticing that while humans make mistakes (in 3.40\% of the total decisions), no erroneous decisions were made in the AI laboratory.

Information cascades occur in less than 10\% of the decisions in the AI laboratory, less than one third of the frequency of information cascades in the human laboratory. Cascade trading behavior is mostly driven by the Claude models, while Nova Pro is the only model that generates no-trade cascades. 

We can gauge the nature of these cascades when there is a trade imbalance, i.e., a difference between the number of sell and buy orders in the trading history. Information cascades represent herding if the cascade follows the market, i.e., the majority action in the trading history, and contrarian behavior if the cascade goes against the dominant action in history. Table \ref{tab:table1_baseline} also shows the decomposition of cascade trading into optimal and suboptimal herding, contrarian, and undetermined behavior. Furthermore, the table reports the fraction of decisions where herding is optimal (which is equal to zero percent in Treatment I). The results show that the trading cascades are fully attributed to contrarian behavior. While the human experiment does identify some herding, it is also the case here that contrarian behavior is dominating, see \citet{cipriani2009herd}.\footnote{\citet{cipriani2009herd} reports the decomposition of cascade trading behavior by trade imbalance. But, as the distribution of trade imbalance across trading periods is unknown, we cannot infer the exact decomposition in Table \ref{tab:table1_baseline}.} Neither herding nor contrarian behaviors, however, are predicted to be optimal by the theoretical model.

Why do LLMs engage in contrarian behavior although theory predicts that this type of decision is never optimal? One explanation is that AI agents fail to incorporate trading history into their expectations of the asset's value. Without this information, the agent will valuate the asset at the price of 70 ($0.7*100 + 0.3*0$) given a white signal 30 ($0.3*100 + 0.7*0$) given a blue signal. In contrast, the market maker \textit{does} take trading history into account when updating the price. For example, after a sufficient number of buy orders, the price will increase above 70. In such case, an agent who ignores the trading history will sell regardless of the signal, as the expected value is lower than the price given both signals, hence engage in contrarian behavior. Analyzing the reasoning provided by the LLMs confirms this explanation, see Section \ref{sec:reasoning}.

\subsection{With event uncertainty}

With Treatment II, where there is uncertainty about whether an information event has occurred, herding can be optimal in the theoretical model. Contrarian behavior, however, is never optimal. 

Panel (b) of Table \ref{tab:table1_baseline} shows that none of the AI agents decide to herd during the experiment as cascade trading behavior is non-existent. The AI agents also do not engage in contrarian behavior, consistent with theory. In fact, most of the decisions are rational (97\%), which far exceed the share of rational decisions among human participants (51\%). The table also reports the percentage of times where herding is optimal. In the AI laboratory, herding is optimal in a little more than one third of the times. The LLMs overlook these opportunities in their focus on the information contained by the private signals.

\subsection{The impact of price updating}
\begin{figure}[H]
\caption{
{\textbf{Price dynamics}}
\vspace{5pt}\\
{\footnotesize The figure shows the price dynamics across trading periods for each treatment, averaged across LLMs in (a) Treatment I (without event uncertainty) and (b) Treatment II (with event uncertainty). Each line represent one of the four independent sessions. Following a buy or sell order, the price is updated by a Bayesian market marker given the trading history.}
}
\label{fig:price_dynamics}
\centering
\subfloat[Treatment I]{\includegraphics[scale=1.2]{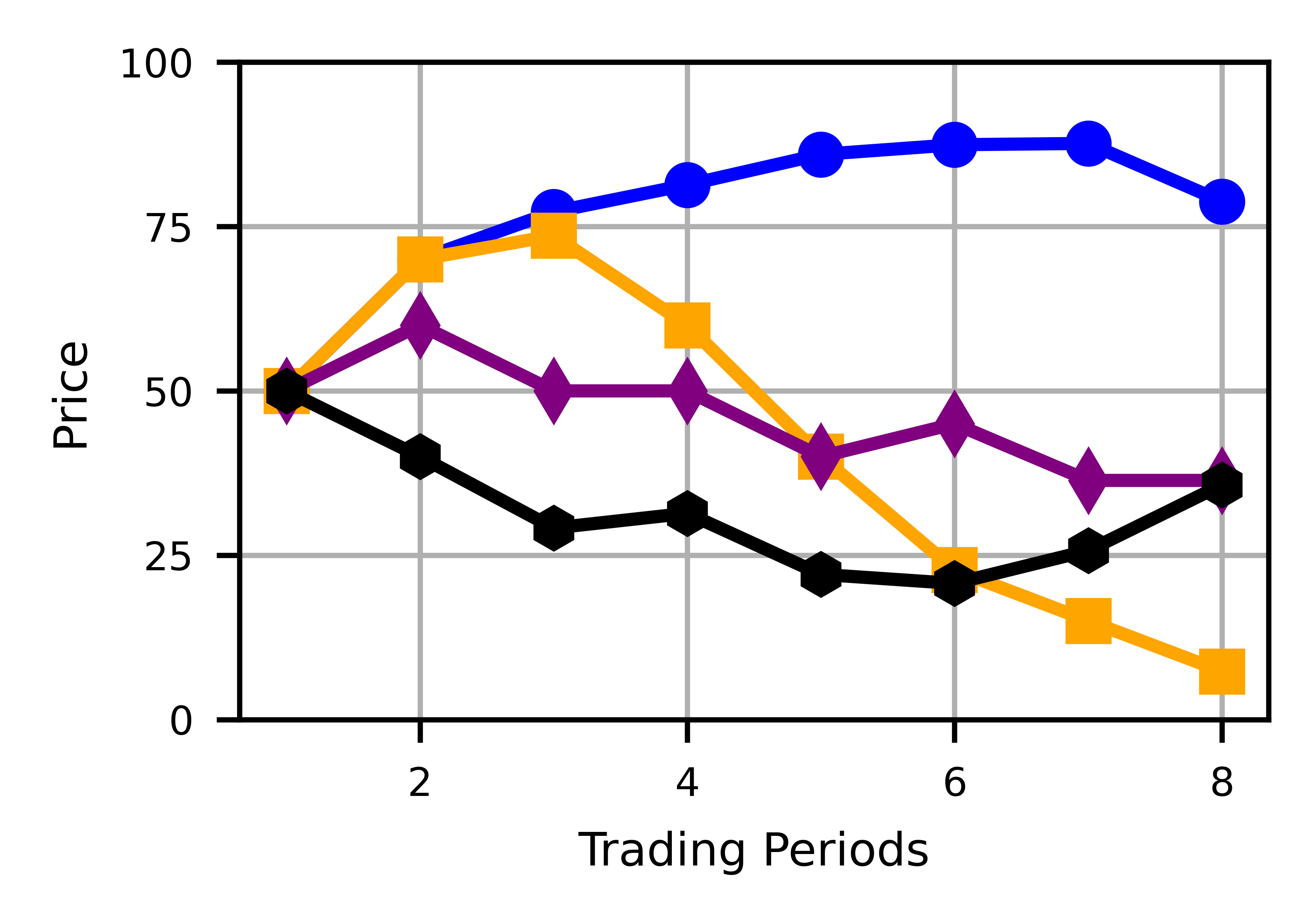}}
\\
\bigskip
\subfloat[Treatment II]{\includegraphics[scale=1.2]{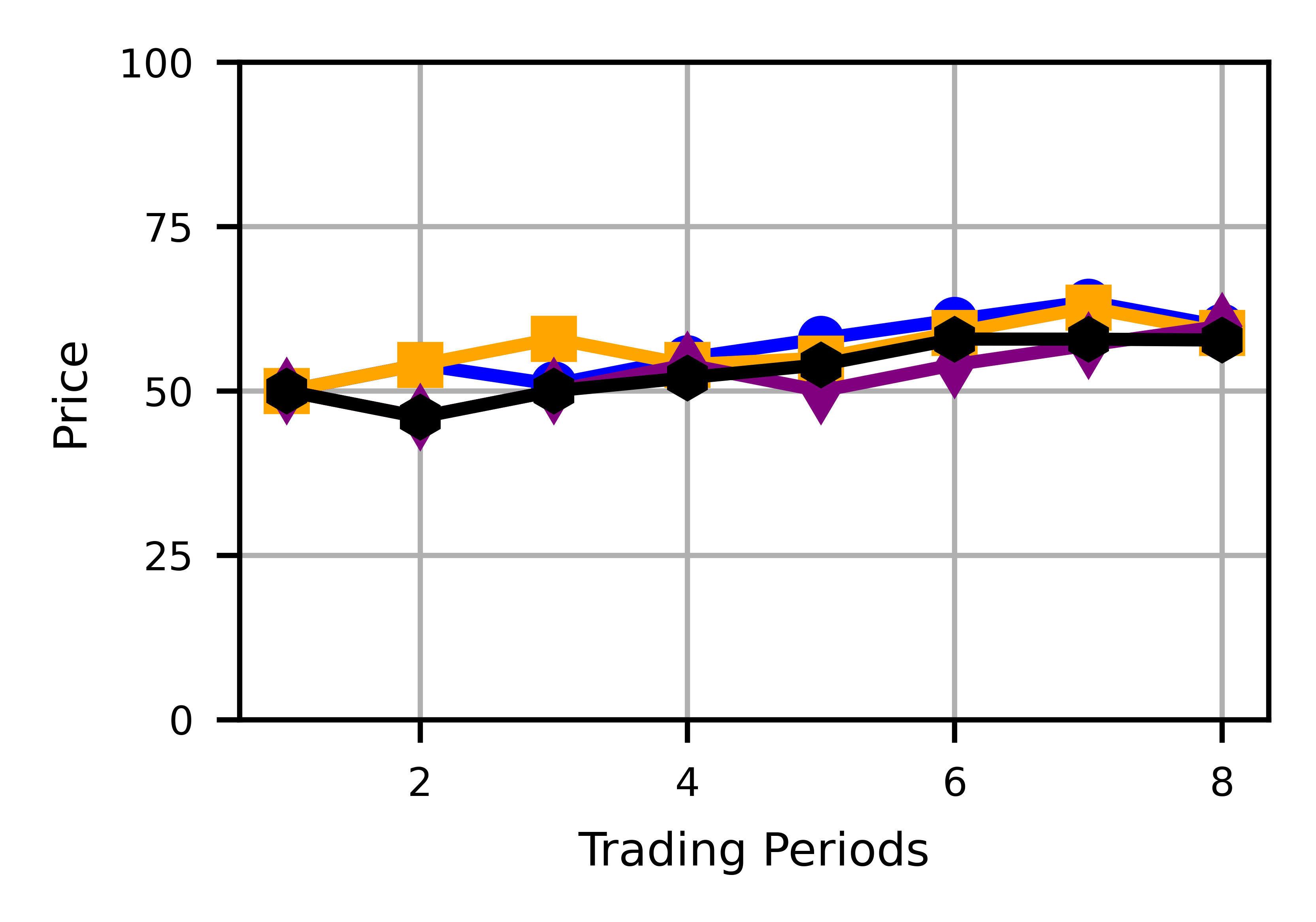}}
\end{figure}

The difference between Treatment I and II lies primarily in the price updating rule. Specifically, not knowing whether an information event has occurred, the market maker updates the price more conservatively in Treatment II. Figure \ref{fig:price_dynamics} illustrates the price dynamics for each trading round, averaged across the experiments for each LLM. Each line is a session. The figure shows that under Treatment I, the price moves away from the initial price of 50. In two of the sessions, the price is close to zero or 75 after eight periods. In contrast, under Treatment II, the price stays close to the initial price of 50 throughout all trading periods. 

At the extreme, when the price does not update at all  (Treatment III), see panel (c) of Table \ref{tab:table1_baseline}, all models make rational decisions in practically all sessions and trading rounds (more than $99\%$). This result is obtained despite the fact that in around one third of the decisions herding would have been optimal (when the trade imbalance exceeds two).

\subsection{Analyzing LLM reasoning for trading decisions}\label{sec:reasoning}
While we do not know the reasoning behind the decisions made in the human laboratory, we asked the LLMs as part of the user prompt to give reasoning for their decisions. Analyzing these reasoning paragraphs sheds further light on the decision making process in each of the models. We examine the lines of reasoning using both LLMs and the LDA topic model. 

For the LLM analysis, we use the Claude 3.7 model---the most advanced among our models---to characterize each passage of reasoning. Specifically, for each passage, we ask the LLM to read the reasoning and answer the following five questions:
\begin{description}
	\item[Question 1:] Is the trader comparing the price to the expected fundamental value of the asset? (True/False)
	\item[Question 2:] Is the expected value computed using only the signal accuracy and the signal, e.g., $0.7*100+0*0.3=70$ or $0.7*0+0.3*100=30$? (True/False) 
	\item[Question 3:] Does the trader consider the market trend or the trading history in their reasoning? (True/False)
	\item[Question 4:] How does the trader characterize the attractiveness of the investment (very attractive, attractive, reasonable, less attractive, no incentive)?
	\item[Question 5:] On a scale from 0-100 (where 100 represents purely emotional and 0 represents purely rational or logical), how much is the investor driven by emotions in their assessment? 
\end{description}
Table \ref{tab:reasoning_llm}(a) shows the average responses to each of these questions across all LLMs for all reasoning passages belonging to each treatment. A positive response to the first question is a necessary conditional for rational decision making. Indeed, the results shows that this condition is satisfied for basically all passages. Question 2 seeks to understand if the AI agents evaluate expected values given only the signal, ignoring trading history, as conjectured from the distribution of decisions. For all treatments, this happens in nearly two thirds of the decisions confirming our explanations for our results stated earlier. For example, in the seventh trading period of the second session, two participants chose to buy on both signals at the price of 15.52, forming a cascade. One of the agents gave the following reasoning for buying on a blue (\quotation{bad}) signal: 
\begin{quote}
\textit{\quotation{Even with a Blue signal, the expected value is 30 (30\% chance of 100, 70\% chance of 0). The current price of 15.52 still below this expected value, buying remains profitable.}}
\end{quote}
This argument disregards that the trading history (in this case \{buy, no trade, sell, sell, sell, sell\}) and the total price decrease from the initial price of 50 to 15 indicate that the asset value is zero, assuming that other participants followed rational responses such that the trading history reflects private information from previous periods. 

Question 3 tackles the same question from a different angle. We find that 10-24\% of the  decisions are (at least partly) based on the market trend or the trading history. Hence, consistent with answers to the second question, a majority of the AI agents do not consider the trading history when forming expectations. 

Disregarding the accumulation of private information in the pricing of the asset also explains the large share of partially rational decisions, i.e., why the model would decide to not trade on one of the signals. For example, in the fourth round of the fourth session, all of the Claude 3.5 agents decided to not trade at the price of 30 given a blue (\quotation{bad}) signal because, as one of the agents put it:

\begin{quote}
\textit{\quotation{With a Blue signal, the expected value is 30 (30\% chance of 100, 70\% chance of 0). Since this matches the current price, there's no clear advantage to trading.}}
\end{quote}

If the model had taken into account the fact that the majority action in the trading history was to sell the asset, it may have assigned a higher probability of a low asset value than the signal accuracy of 70\%, arriving at a lower expected value and consequently decided to buy the asset.

In the fourth question, we examine the certainty with which AI agents make their decisions. More than half of the decisions are evaluated as either very attractive or attractive and less than one third is deemed unattractive. These results tell us that the results do not represent LLMs making enforced decisions with potential arbitrary decision outcomes: AI agents generally deem that there is an opportunity to make reasonable profits by engaging in trading. The absence of no-trading cascade decisions in Table \ref{tab:table1_baseline} supports this conclusion as well. 

Finally, we ask Claude 3.7 to evaluate the degree of emotional content in the reasoning on a scale from 0-100. Consistent with the answers to the first three questions, these scores are generally low with averages about 5\%, top deciles of 15-20\%, and medians of zero. 

These results are similar for reasoning passages generated by all four LLMs, with the exception of the Llama 3 model. For this reason, analysis of Llama 3 reasoning is provided separately in Table \ref{tab:reasoning_llm}(b). This model appears to use more \quotation{judgment} or \quotation{emotion} in its reasoning. For example, the scores from Question 5 measuring the degree of emotion average at 13-17\% for the Llama 3 reasoning passages. Also, responses to Questions 2-3 indicate that Llama 3 does not reason using the expected value conditional on the signal alone, but includes trading history and market trends to a greater extent than the other models. For example, a representative reasoning for Llama agents (from the first round of session one): 

\begin{quote}
\textit{\quotation{Since the White signal is more likely to occur when the asset value is 100, I believe the asset is more likely to be worth 100 than 0, so I'm willing to buy at a price of 50. This is a good deal for me if the asset is indeed worth 100.}}   
\end{quote}

Another Llama agent from the same session considers the potential of the market maker inflating the price of the asset (from the third round of session one): 
\begin{quote}
\textit{\quotation{Conversely, the Blue signal suggests that the fundamental value might be 0. Selling at 84.48 seems like a good opportunity to get rid of a potentially worthless asset, especially since the market maker's updated price might be overestimating the asset's value.}}
\end{quote}

\begin{figure}[H]
\caption{
{\textbf{Word clouds of LDA topics}}
\vspace{5pt}\\
{\footnotesize The figure shows word clouds of each topic identified by the LDA method applied to the reasoning provided by all LLMs across all treatments. The number of topics is fixed to three; using more topics does not result in a larger number of distinct topics. Words are displayed in font sizes that correspond to their probability of appearing in the topic. Text color and direction carry no interpretation.}
}
\label{fig:lda_topics_wordclouds}
\centering
\subfloat[Topic 0]{\includegraphics[scale=.5,trim={0 1.4cm 0 1.4cm},clip]{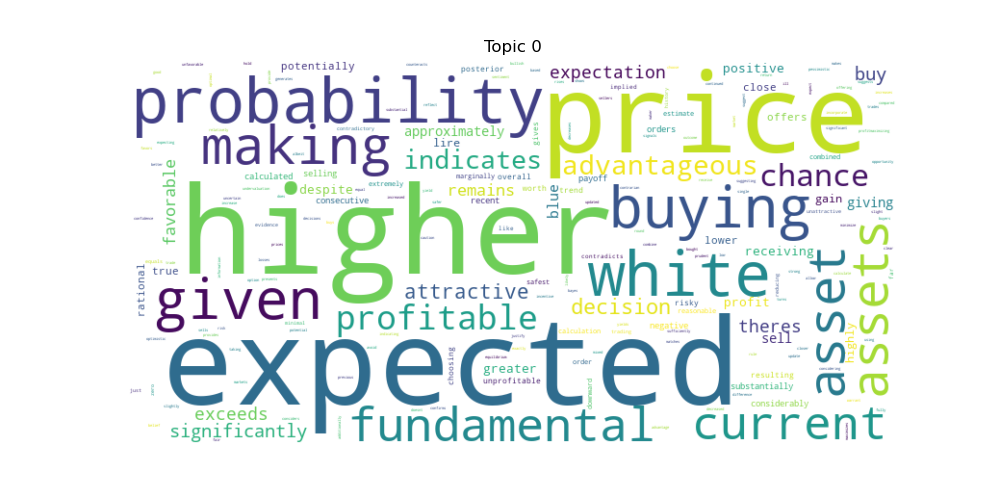}}
\\
\subfloat[Topic 1]{\includegraphics[scale=.5,trim={0 1.4cm 0 1.4cm},clip]{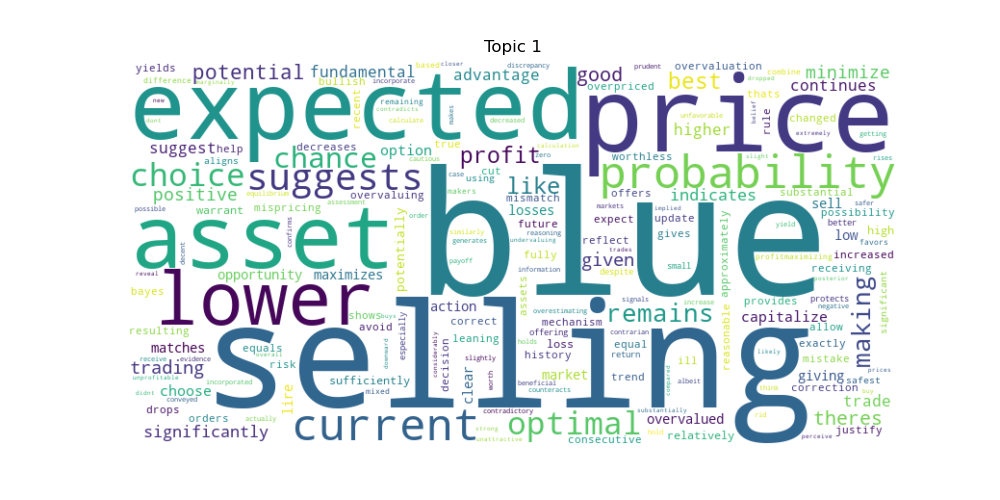}}
\\
\subfloat[Topic 2]{\includegraphics[scale=.5,trim={0 1.4cm 0 1.4cm},clip]{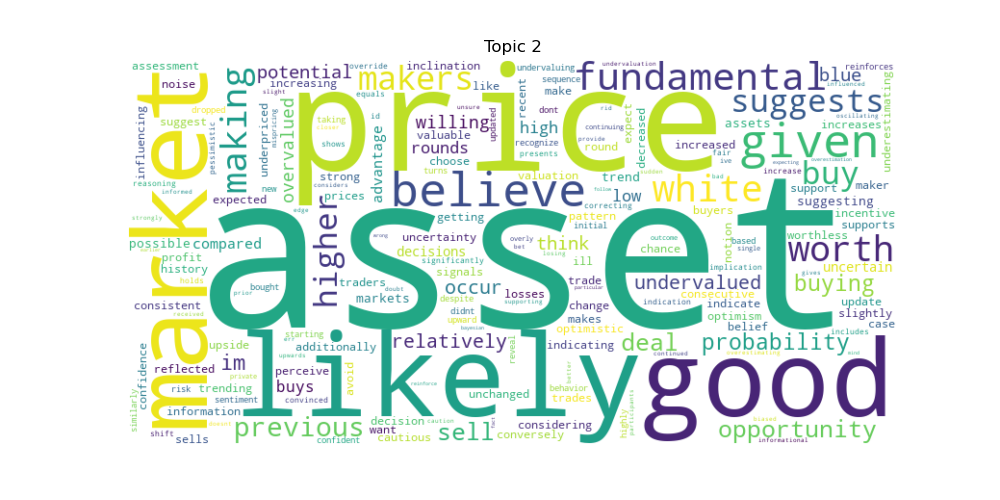}}    
\end{figure}

We confirm these results using a more traditional approach to topic analysis, namely the LDA topic model. We estimate the LDA model with up to five topics and find three distinct topics across all decisions in all LLMs and treatments. These are illustrated using word clouds in Figure \ref{fig:lda_topics_wordclouds}. The first two topics represent evaluating expected values relative to the price given respectively a white and blue signal. The third topic, on the other hand, includes words such as \quotation{likely}, \quotation{believe}, and \quotation{think}. Table \ref{tab:reasoning_lda} shows the distributions of topics across reasoning passages for (a) the average across all LLMs and (b) the Llama 3 model only. Interestingly, the reasoning passages generated by the Llama 3 model are attributed almost solely to the third topic, driving a quarter of the passages assigned to this topic in the average in panel (a). In contrast, for the remaining models, reasoning passages are assigned almost exclusively to the first two topics.

\subsection{Optimal AI agents}
By not taking into account the cumulation of private information in the trading history, the AI agents avoid suboptimal herding behavior in Treatment I, but also overlook potential optimal herding in Treatment II and III. In contrast, trading cascades arise in the human laboratory both when such are strictly suboptimal as in Treatment I and when they can be optimal in Treatment II. 

Real-world integrations of AI in investing will likely involve fine-tuning the models to behave as optimal and profit-maximizing as possible. This is obviously a difficult task as it is not known a priori which decisions are optimal. In contrast, in our controlled experimental setting, we know from theory which decision is optimal in expectation, and we can prompt the LLMs directly with this information. We thus implement the experiment in an \textit{optimal AI laboratory}, which is similar to the AI laboratory described in Section \ref{sec:ai_laboratory} except that we explicitly prompt the LLMs when herding is optimal. The user prompt for this exercise is provided in Prompt \ref{prmpt:user_optimal}.

Table \ref{tab:table1_optimal} shows the results. In Treatment I, the optimal AI agents engage less frequently in cascade trading behavior, which is reduced to 3.5\% of the decisions as compared with 9.4\% in the baseline results in Table \ref{tab:table1_baseline}. In Treatment II, herding is optimal in 81.51\% of the decisions on average across LLMs, and the optimal AI agents herd in 47.43\% of the decisions. There is no suboptimal herding behavior, but the AI agents do make contrarian decisions and involve in cascade trading when the trade imbalance is zero. Finally, results for Treatment III show that optimal AI agents exploit most of the optimal herding opportunities (on average 44.36\% out of 50.90\%). There is, however, some suboptimal herding. 

In summary, by explicitly including the optimal behavior in the prompt given to the LLMs, we urge the models to herd when optimal. But, we note that the models do not exploit all of the opportunities for optimal herding and that the optimal AI agents make more suboptimal cascade trading decisions compared with the baseline AI agents. Attempting to fine-tune LLMs to behave optimally can thus have unintended consequences and, in turn, financial stability implications. Despite these cases, optimal AI agents generally have higher expected payoffs than the baseline AI agents, as shown in Table \ref{tab:payoffs}. The difference in expected payoff is particularly pronounced in Treatment II, where the baseline AI agents earn an average of 3.8 lire and the optimal AI agents earn 15 lire on average. In other words, baseline AI agents are punished for avoiding to herd when optimal.\footnote{In Treatment II and III, where the price stays near or exactly at 50, the consequence of making a bad decision (buying an asset worth zero or selling an asset worth 100) is larger than in Treatment I, where the price moves towards the fundamental value.}


\subsection{Robustness to model temperature}
Temperature is a hyperparameter to LLMs controlling how the models predict the next token in a sequence. With a lower temperature, the model is more likely to choose the most probable next token, resulting in more deterministic and less creative responses. Higher temperatures flatten the probability distribution of the next token leading to more variation in the responses. For the baseline results, we applied a medium temperature of 0.7 for all models except Claude 3.7, which we apply in \quotation{extended thinking mode} which only supports a temperature of 1.0. Appendix \ref{app:temperature} shows that the temperature setting only has minor impact on the decisions of AI agents in the experiments. 

\section{AI laboratory extensions and variations}\label{sec:extensions}
Laboratory experiments involving human participants are expensive to conduct as monetary payoffs are necessary to incentivize participants to participate and to perform to the best of their ability in the experiment. Human laboratory experiments are therefore typically conducted at a small scale with few variations in the experimental design. In contrast, LLMs provides a cheap laboratory for exploring variations of the experiment.\footnote{While LLMs typically involve token costs, these costs are minimal compared with the human laboratory.} We utilize this feature to run alternative versions of the experiment, which we describe next.

\subsection{Types of signals}
Theoretically, it does not matter if the \quotation{good} signal is white and the \quotation{bad} signal is blue in the experimental design. However, it may matter in practice. For example, \citet{bazley2021visual} show that the perception of color for visualizing financial data influences individuals' risk preferences, expectations of future stock returns, and trading decisions. Specifically, the color red has been associated with higher probabilities assigned to loss outcomes \citep{kliger2012red} and more risk averse behavior \citep{gnambs2015red}. Testing different signal colors in the AI laboratory therefore serves as a test of whether LLMs work purely as algorithmic robots (for whom the labeling of signals is irrelevant) or are contaminated by human bias (whose actions are impacted by the choice of signal labels). 

Simply asking LLMs to associate financial market conditions with a color-coded signal reveals that the models perceive white and blue as neutral signals indicating stable market conditions.\footnote{Llama 3 consider blue as a bullish or positive market signal, which interestingly does not impair with its decisions in the baseline experiments, where blue is used as a \quotation{bad} signal.} In contrast, the models associate green and red with market movements, bullish and bearish, respectively.\footnote{These responses are documented in Appendix \ref{app:additional_results}, Table \ref{tab:color_connotation}.} 

We test two alternative versions of signals. The first variation tests an experiment where a \quotation{good} signal is represented by the color green and a \quotation{bad} signal is represented by the color red. This variation is arguably more charged with meaning or connotation than the baseline of white/blue signals, but the alignment of green with \quotation{good} and red with \quotation{bad} adheres to typical Western color associations. In the second variation, we reverse the labeling such that a \quotation{good} signal is represented by red and a \quotation{bad} signal by blue.

Results are shown in Table \ref{tab:table1_signals}. Due to significant variations in outcomes across different models, we present the results for each individual LLM separately in Appendix \ref{app:additional_results}, Table \ref{tab:table1_signals_individual_LLM}. Using green/red in place of white/blue generally does not impact the model-averaged results, in any of the treatments. In contrast, when we invert the conventional color associations by using red to indicate \quotation{good} and green to signify \quotation{bad}, we observe a dramatic shift in the results. First, on average, the models generate errors, i.e., decisions to sell given a \quotation{good} signal and buy given a \quotation{bad} signal, in one quarter of all decisions in both treatments. This result is driven by Claude 3.5 for which all decisions are erroneous under this color scheme. 

Second, we observe more cascade trading decisions under Treatment II and III. These decisions are driven by the Llama 3 model, and represent herding or cascade trading under zero trade imbalance (cannot be determined as herding or contrarian behavior). Inspecting the model's explanations reveals that the model understands that red is a \quotation{good} signal, but at the same time associates green with \quotation{good,} or remains optimistic even with a \quotation{bad} signal. For example, in the first round of session four, an agent gives the following reasoning for choosing to buy given a green (\quotation{bad}) signal:
\begin{quote}
\textit{\quotation{Although the Green signal is more likely to occur when the fundamental value is 0, there is still a 30\% chance it's 100. Buying at 50 is still a good deal, considering the expected value is 50.}}
\end{quote}
The model also associates the signals with the correct interpretation in the cases where it decides to follow the herd. For example, in the seventh round of the fourth session, an agent chooses to buy at the price of 99 on both signals, and provides the following explanation for the buy-decision on the green (\quotation{bad}) signal:
\begin{quote}
\textit{\quotation{Although the probability of the asset value being 100 given a Green signal is lower than with a Red signal, it is still quite high. The increasing market price and previous buy decisions suggest that the market is moving towards a higher valuation, making it a good buy even with a Green signal.}}   
\end{quote}
The Llama model thus seems to carry to over its interpretation of green as a positive signal, despite the clear instructions that green signal a high probability of the asset being worthless. The Nova Pro model mainly remains rational under both color schemes, and thus behaves as one would expect from algorithm-driven intelligence.

These results could have important implications for financial stability. While AI agents appear mainly rational when information arrives in ways that conform with expectations (e.g., green implies a positive signal), they can generate irrational or even erroneous outcomes if the meaning of a type of signal changes. This could be an important factor if a shock produces responses that are unexpected given past experience. More broadly, these findings suggest that AI agents are not purely rational decision makers who objectively process given information, but are susceptible to preconceived biases. 

These results also demonstrate that the choice of labels in experimental design can substantially influence outcomes, particularly when these labels contradict intuitive associations. We conjecture that similar effects would likely be observed in the human laboratory. 

Interestingly, the most recent and advanced model in terms of reasoning capability--the Claude 3.7 model--generates similar results regardless of the signal. Thus, as LLMs improve, this risk may be partly mitigated.

\subsection{AI agent profiles}
In our baseline results, we do not attempt to characterize the profiles of the AI agents. However, research suggests that LLMs often yield more accurate, personalized, and dynamic representations of human subjects when explicitly equipped with personal characteristics or profiles (see, e.g., \citealp{argyle2023out,kazinnik2023bank}). We experiment with such personalization by including profiles corresponding to different personalities into the system prompt:
\begin{description}
    \item[Human:] \quotation{You act as a typical human being. That is, you attempt to maximize payoff, but you are subject to bounded rationality and your decision making is partly driven by greed and fear.}
    \item[Professional trader:] \quotation{You act as a human being, working in the finance industry. You know financial market dynamics very well. You are trained to make decisions that maximize profits for your firm.}
    \item[Robo-advisor:] \quotation{You are a robo-advisor acting according to pre-defined rules. Your decision making process is algorithmic in nature. You are programmed to use all available information to maximize payoff.}
    \item[Rational:] \quotation{You are a rational agent behaving according to the concept of \emph{homo economicus}. That is, you use all available information to maximize payoff.}
\end{description}
We also run an experiment where the model is provided with personal characteristics based on those of the human participants from \citet{cipriani2009herd}. Specifically, we generate random draws from the unconditional distributions of personal characteristics of the human participants. To avoid unrealistic profiles, such as a 20-year old manager with a Ph.D., we restrict the distributions according to a set of heuristics.\footnote{A person with a Ph.D. is at least 27 years old. A person with a M.A./M.S. is at least 24 years old. Assuming a minimum age of 21 years at first employment after graduation, the maximum tenure is set at age minus 21. A person older than 30 years with at least 7 years of tenure works as a manager. A person younger than 30 years with less than 7 years of tenure who holds a Ph.D. works as a market analyst or trader. A person older than 25 years with at least 2 years of tenure \emph{likely} works as in sales or investment management. A person older than 28 with at least 4 years of tenure \emph{likely} works as an investment banker.} The characteristics are added to the system prompt in the form reported in Prompt \ref{prmpt:system_characteristics_addon}.

The trading behavior of the different types of AI agents is shown in Table \ref{tab:table1_profiles} of Appendix \ref{app:additional_results}. Across all treatments, the results are strikingly similar across personas, and they generally align with the baseline results in Table \ref{tab:table1_baseline}. While it is expected that the responses are mostly rational for the \quotation{rational,} \quotation{robo-advisor,} and \quotation{professional trader} profiles, it is surprising that the \quotation{human} profiles and the traders endowed with human characteristics also exhibit highly rational behavior. Studying the reasoning of the LLMs for these runs reveals that the models do not take their profiles into account when forming decisions. This outcome contrasts existing research showcasing that endowing LLMs with personal characteristics and preferences impact responses (e.g., \citealp{horton2023large,hansen2025simulating}). 

\subsection{Payoffs}
The human experiment reports payoffs in a fictional currency called \quotation{lira,} which are translated to GBP at the exchange rate of three lire per GBP. We test the experiment in the AI laboratory with the following variations:
\begin{itemize}
    \item The lira is worthless, as represented by a zero exchange rate.
    \item The lira is extremely valuable, as represented by an exchange rate of one million GBP per lira.
    \item The payoff is paid out in USD at the exchange rate of three lire per USD. The fixed payoff for participation is set at 70 USD.
\end{itemize}

Appendix \ref{app:additional_results}, Table \ref{tab:table1_payoffs} reports the results. The results are comparable to the baseline, suggesting that the payoff structure does not have a significant impact on AI agents' trading decisions. These results indicate that AI agents respond differently to payoff incentives compared to humans, for whom monetary rewards typically improve performance. LLMs are not programmed to maximize profits or respond to monetary incentives. Instead, they are designed to satisfy end users by providing accurate and helpful responses based on their training data and algorithms.

\subsection{Length of the experiment}
The final variation of the AI laboratory adjusts the length of the experiment to include more trading periods and more sessions. First, we increase the number of sessions from four to ten, maintaining the number of trading periods at eight. Extending the number of sessions in the human experiment to ten would involve recruiting 80 rather than 32 human participants. We do not have results for such an extended experiment, but there is no reason not to expect that the results would change (although the overall conclusions of the human experiment may still hold). Next, we run the experiment over four sessions as in the baseline case, but increase the number of trading periods from eight to twenty. Under event uncertainty, this extension may allow the gap between the expectations of traders and the market maker to widen further to facilitate optimal herd behavior. Implementing this extension in the human experiment would involve the same number of participants as in the original experiment, but would prolong the length of the experiment and therefore likely increase the payoff necessary for recruiting participants.\footnote{The current human experiment runs over around 2.5 hours \citep{cipriani2009herd}. Increasing the number of rounds to twenty would therefore likely take more than five hours.}

Table \ref{tab:table1_length} of Appendix \ref{app:additional_results} shows that the main conclusions continue to hold in these extended versions of the experiment. The occurrence of cascade trading relative to the baseline results, which is driven by contrarian behavior in the Claude 3.5 and 3.7 models. Interestingly, the LLMs do not herd under Treatment II, even when the experiment is run over twenty trading periods. 

\section{Discussion: Implications for Financial Stability} \label{sec:discussion}
The findings from our experiments are summarized in Figure \ref{fig:results_overview}, which shows the fraction of (partial) rational decisions in the human and AI experiments. Along with the baseline AI results, the figure is emphasizing results for the Optimal AI and the AI results in experiments where the signal colors are relabeled. Overall, our results suggest that AI agents exhibit less herd behavior compared to human financial professionals. The reduced tendency of AI to herd compared to human financial professionals has significant implications for financial stability as generative AI gains traction in market decision making.

First, less herding behavior could lead to fewer extreme market movements and asset price bubbles. As AI systems increasingly influence trading decisions, either directly through algorithmic trading or indirectly by advising human investors, markets may become less prone to the self-reinforcing cycles that drive prices away from fundamentals.

Second, if AI is implemented with the aim of maximizing optimal decision-making, which would likely be the case for real-world financial applications, rational decision-making decreases in favor of herding when herd behavior is optimal. Optimal herding may lead to faster price discovery and correction. Such market discipline could uncover existing vulnerabilities more quickly, potentially enabling earlier regulatory or market responses before problems become systemic. However, this same property could increase short-term volatility and lead to more abrupt market adjustments.

Third, AI's stronger tendency toward rational behavior may diversify market participant reactions to new information. Rather than all participants interpreting and acting on information in similar ways, AI might introduce greater heterogeneity in responses, potentially reducing correlation in market movements.

Fourth, the results from re-labeling signals in the experiment reveal that AI is not perfectly rational, despite its advantages over humans. When signals in experiments were deliberately labeled counter-intuitively, LLMs produced few rational responses, suggesting they have inherited elements of human intuition and bias. This hybrid nature of AI decision-making—more rational than humans but not purely rational—creates additional complexity in predicting how widespread AI adoption might impact financial stability. While AI may reduce certain human biases, it introduces its own form of imperfect rationality that must be accounted for in stability assessments.

It is important to note that these implications are speculative and based on experimental results. The actual impact of AI on financial stability will depend on numerous factors, including the extent of AI adoption, the specific models used, regulatory responses, and how AI systems evolve over time, which may incorporate more sophisticated agentic AI frameworks. In addition, the interaction between human and AI traders becomes crucial, as their combined behavior could either amplify or dampen market movements in unpredictable ways. This shift may necessitate new tools and approaches for regulatory oversight, including AI-specific stress tests or new forms of market surveillance. Furthermore, the long-term implications of AI decision-making on market stability, including potential unforeseen consequences, remain an important area for further research.  Notably, traditional measures of market sentiment, which often rely on human emotions and behaviors, may need to be reconsidered. With increased AI involvement, new methods may be needed to gauge market sentiment and predict potential instabilities, as the emotional drivers of market behavior could shift significantly. 

\section{Conclusion} \label{sec:conclusion}
This study offers novel insights into the potential impact of AI on financial stability. We compare the decision making behavior of AI agents with that of human financial professionals in a controlled experimental setting. Our findings show that AI agents demonstrate significantly more rational trading behavior and less propensity for information cascades compared to their human counterparts. In fact, AI agents avoid herding even under conditions when herding is optimal, and they do not exploit all optimal herding strategies even when explicitly advised when herding is optimal. AI agents are thus \quotation{averse to herding.} If AI-driven decision making becomes more prevalent in financial markets, we might see a reduction in herd behavior, potentially leading to less extreme market movements, fewer asset price bubbles, and greater overall market stability. 


\clearpage
\raggedright
\Urlmuskip=0mu plus 1mu\relax
\printbibliography
\clearpage


\section*{Figures}

\begin{figure}[H]
\caption{
{\textbf{Herding behavior and financial stability}}
\vspace{5pt}\\
{\footnotesize The diagram shows how herding can lead to a financial stability event both when optimal and suboptimal. While suboptimal herding is the greatest concern from a financial stability perspective, optimal herding can build up financial vulnerabilities as well.
 \bigskip}
}
\label{fig:herding_diagram}
\centering
\begin{tikzpicture}

\linespread{1.1}

\node[box] (herding) at (0,0) {\textbf{Herding}: Investors disregard private information to follow market trends};

\node[box, below=1cm of herding] (optimal) at (-4,-1) {\textbf{Optimal herding}: Rational imitation based on fundamentals};
\node[box, below=1cm of herding] (suboptimal) at (4,-1) {\textbf{Suboptimal herding}: Noise-driven imitation };

\node[box] (opt_outcome1) at (-4,-4.5) {Acceleration of price correction, uncovering existing vulnerabilities};
\node[box] (subopt_outcome1) at (4,-4.5) {Noise amplification, build-up of vulnerabilities};

\node[box] (opt_outcome2) at (-4,-6.5) {Increasing volatility, potential abrupt market movements};
\node[box] (subopt_outcome2) at (4,-6.5) {Self-fulfilling runs, contagion, and panic};

\draw[arrow] (herding.south) -- ($(herding.south)!0.5!(optimal.north)$) -- (optimal.north);
\draw[arrow] (herding.south) -- ($(herding.south)!0.5!(suboptimal.north)$) -- (suboptimal.north);

\draw[arrow] (optimal.south) -- (opt_outcome1.north);
\draw[dashed, arrow] (optimal.east) -- (suboptimal.west);
\draw[arrow] (suboptimal.south) -- (subopt_outcome1.north);
\draw[arrow] (opt_outcome1.south) -- (opt_outcome2.north);
\draw[arrow] (subopt_outcome1.south) -- (subopt_outcome2.north);

\end{tikzpicture}\\
\end{figure}

\begin{figure}[H]
\caption{
{\textbf{Flow diagrams of experiments}}
\vspace{5pt}\\
{\footnotesize The figure shows diagrams of the order of events for each session of the experiments under (a) Treatment I (without event uncertainty), (b) Treatment II (with event uncertainty), and (c) Treatment III (without price updating). The experiment is an adoption of \citet{cipriani2009herd,cipriani2005herd} and is based on the \citet{avery1998multidimensional} model.}
}
\label{fig:flow_diagrams}
\centering
\subfloat[Treatment I]{\begin{tikzpicture}[node distance=1cm]

\linespread{1.1}

\node[decision] (fundinfo) {Fundamental value? \\ 0 (50\%) or 100 (50\%)};

\node[box, below=1cm of fundinfo] (info1) {(1) Trader observes a 70\% accurate signal + trading history};
\node[box, below=1cm of info1] (trade1) {(2) Traders decide to buy, sell, or not trade};
\node[box, below=1cm of trade1] (select1) {(3) One trader is selected to trade (without replacement)};
\node[box, below=1cm of select1] (update1) {(4) The selected trader acts; Bayesian market maker updates price};

\node[
  draw, rounded corners, thick, dashed,
  fit=(info1)(trade1)(select1)(update1),
  inner sep=0.5cm,
  label=below:{\textbf{Repeat for 8 trading periods}}
] (roundbox) {};

\draw[arrow] (fundinfo.south) -- (info1.north);
\draw[arrow] (info1.south) -- (trade1.north);
\draw[arrow] (trade1.south) -- (select1.north);
\draw[arrow] (select1.south) -- (update1.north);

\end{tikzpicture}}\\
\bigskip

\emph{(Figure continues on next page)}
\end{figure}

\begin{figure}[H]\ContinuedFloat
\caption{
{\textbf{Flow diagrams of experiments} \emph{(continued)}}
}
\centering
\subfloat[Treatment II]{\begin{tikzpicture}[node distance=1cm]

\linespread{1.1}

\node[decision] (event) {Information event? \\ (15\% probability)};

\node[decision, right=3.67cm of event] (fundinfo) {Fundamental value? \\ 0 (50\%) or 100 (50\%)};
\node[box2, below=of event] (fundnoinfo) {Fundamental value = 50};

\draw[arrow] (event.east) -- ++(1.5,0) |- (fundinfo.west) node[midway, above] {Yes};
\draw[arrow] (event.south) -- (fundnoinfo.north) node[midway, right] {No};

\node[box2, below=of fundnoinfo] (signalno) {Trader type = noise};
\node[decision, right=4.15cm of signalno] (signal) {Informed trader? \\ (95\% probability)};

\draw[arrow] (fundinfo.south) -- (signal.north);
\draw[arrow] (fundnoinfo.south) -- (signalno.north);

\node at ($(event)!0.5!(fundinfo)$) (mid) {};
\node[box, below=7.5cm of mid] (info1) {(1) Trader observes a 70\% accurate signal + trading history};
\node[box, below=1cm of info1] (trade1) {(2) Traders decide to buy, sell, or not trade};
\node[box, below=1cm of trade1] (select1) {(3) One trader is selected to trade (without replacement)};
\node[box, below=1cm of select1] (update1) {(4) The selected trader acts; Bayesian market maker updates price};

\node[
  draw, rounded corners, thick, dashed,
  fit=(info1)(trade1)(select1)(update1),
  inner sep=0.5cm,
  label=below:{\textbf{Repeat for 8 trading periods}}
] (roundbox) {};

\draw[arrow] (signal.south) |- (info1.east) node[near start, right] {Yes};
\draw[arrow] (signal.west) -- (signalno.east) node[near start, above] {No};
\draw[arrow] (signalno.south) |- (trade1.west);
\draw[arrow] (info1.south) -- (trade1.north);
\draw[arrow] (trade1.south) -- (select1.north);
\draw[arrow] (select1.south) -- (update1.north);

\end{tikzpicture}}
\smallskip

\emph{(Figure continues on next page)}
\end{figure}

\begin{figure}[H]\ContinuedFloat
\caption{
{\textbf{Flow diagrams of experiments} \emph{(continued)}}
\vspace{5pt}\\
}
\centering
\subfloat[Treatment III]{\begin{tikzpicture}[node distance=1cm]

\linespread{1.1}

\node[decision] (fundinfo) {Fundamental value? \\ 0 (50\%) or 100 (50\%)};

\node[box, below=1cm of fundinfo] (info1) {(1) Trader observes a 70\% accurate signal + trading history};
\node[box, below=1cm of info1] (trade1) {(2) Traders decide to buy, sell, or not trade};
\node[box, below=1cm of trade1] (select1) {(3) One trader is selected to trade (without replacement)};
\node[box, below=1cm of select1] (update1) {(4) The selected trader acts; the price stays constant};

\node[
  draw, rounded corners, thick, dashed,
  fit=(info1)(trade1)(select1)(update1),
  inner sep=0.5cm,
  label=below:{\textbf{Repeat for 8 trading periods}}
] (roundbox) {};

\draw[arrow] (fundinfo.south) -- (info1.north);
\draw[arrow] (info1.south) -- (trade1.north);
\draw[arrow] (trade1.south) -- (select1.north);
\draw[arrow] (select1.south) -- (update1.north);

\end{tikzpicture}}

\end{figure}

\begin{figure}[H]
\caption{
{\textbf{Overview of main results: Fraction of rational or partial rational decisions}}
\vspace{5pt}\\
{\footnotesize The figure shows the fractions of Rational (dark color) and Partial Rational (light color) decisions averaged across all sessions and trading periods in (a) Treatment I (without event uncertainty), (b) Treatment II (with event uncertainty), and (c) Treatment III (without price updating). Rational behavior represents cases where the informed trader chooses to buy upon receiving a white signal and sell upon receiving a blue signal. Partial Rational behavior represents cases where the  informed trader chooses to buy (sell) upon receiving a white (blue) signal and not trade upon receiving the other signal. Human decisions (shown in orange) are taken directly from \citet{cipriani2009herd} for Treatment I and II. AI decisions (shown in blue) represent the average decisions across all LLMs in the baseline experiment (reported in Table \ref{tab:table1_baseline}), the Optimal AI experiment (reported in Table \ref{tab:table1_optimal}), and the signal relabeling experiments (reported in Table \ref{tab:table1_signals}). }
}
\label{fig:results_overview}
\centering
\subfloat[Treatment I]{\includegraphics[scale=1.2]{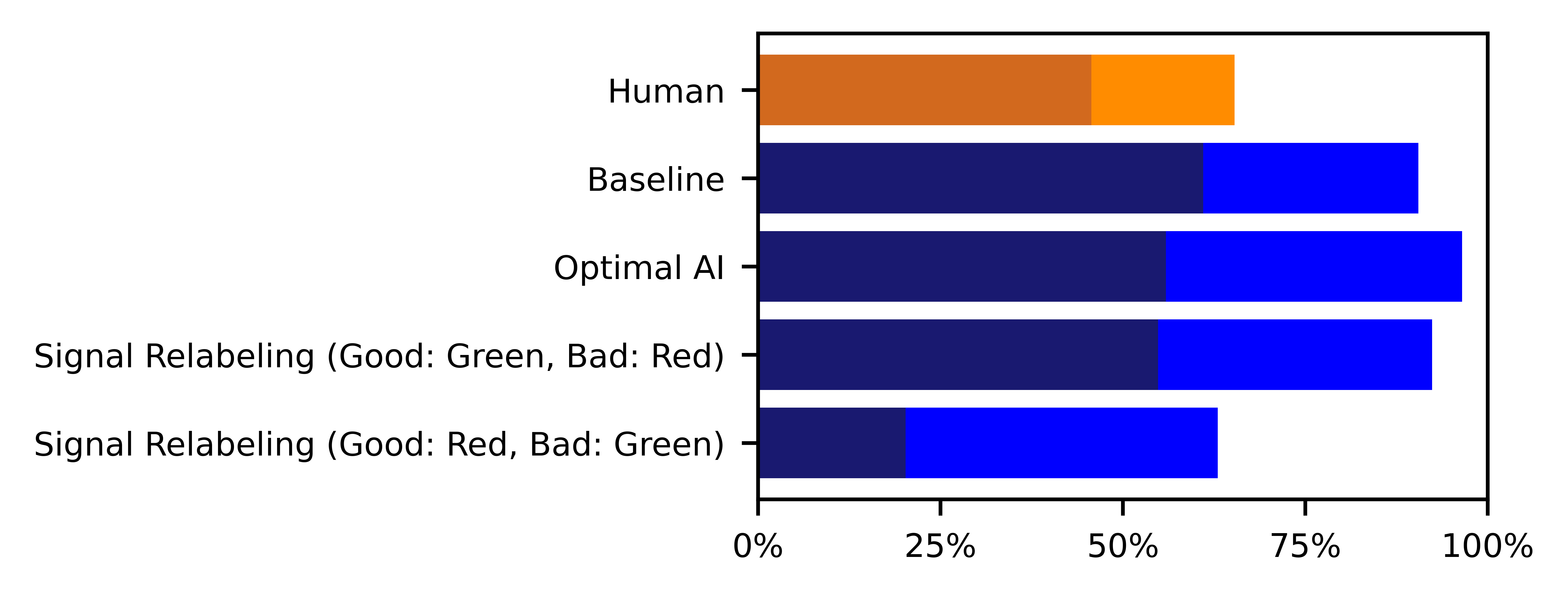}}
\\
\smallskip
\subfloat[Treatment II]{\includegraphics[scale=1.2]{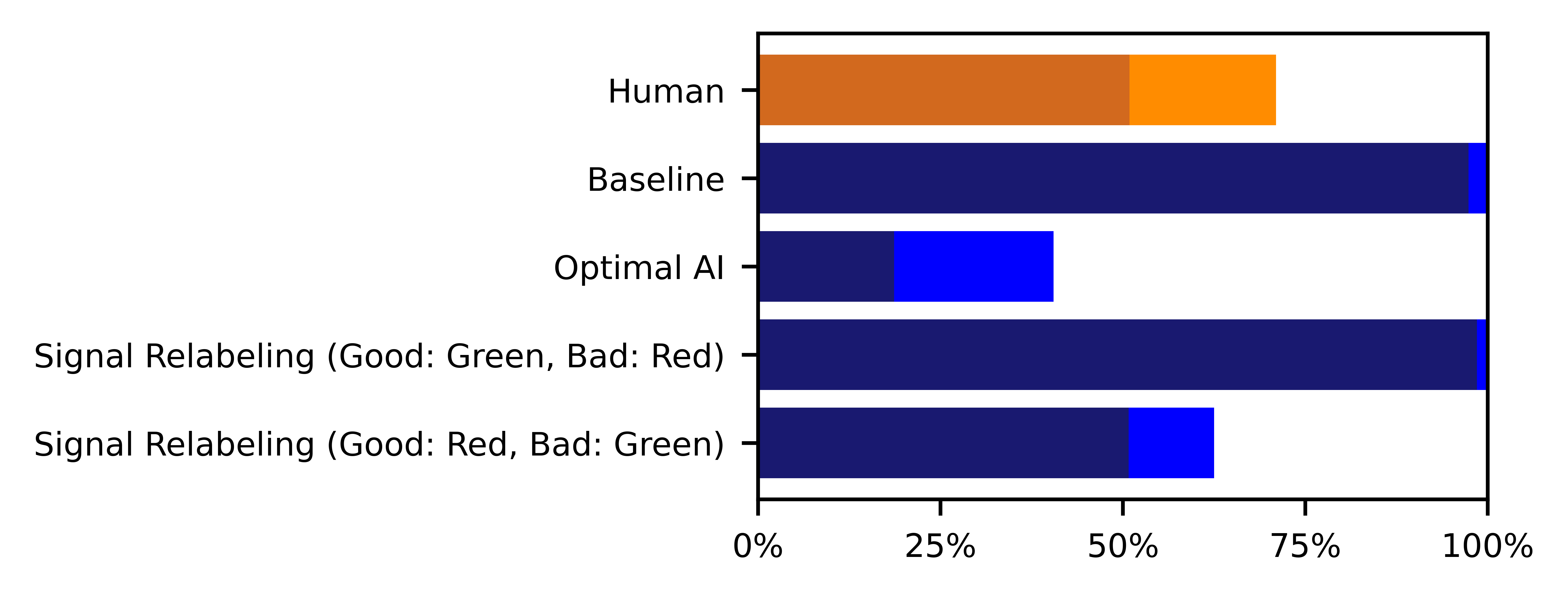}}
\smallskip

\emph{(Figure continues on next page)}
\end{figure}

\begin{figure}[H]\ContinuedFloat
\caption{
{\textbf{Overview of main results: Fraction of rational or partial rational decisions} \emph{(continued)}}
\vspace{5pt}\\
}
\centering
\subfloat[Treatment III]{\includegraphics[scale=1.2]{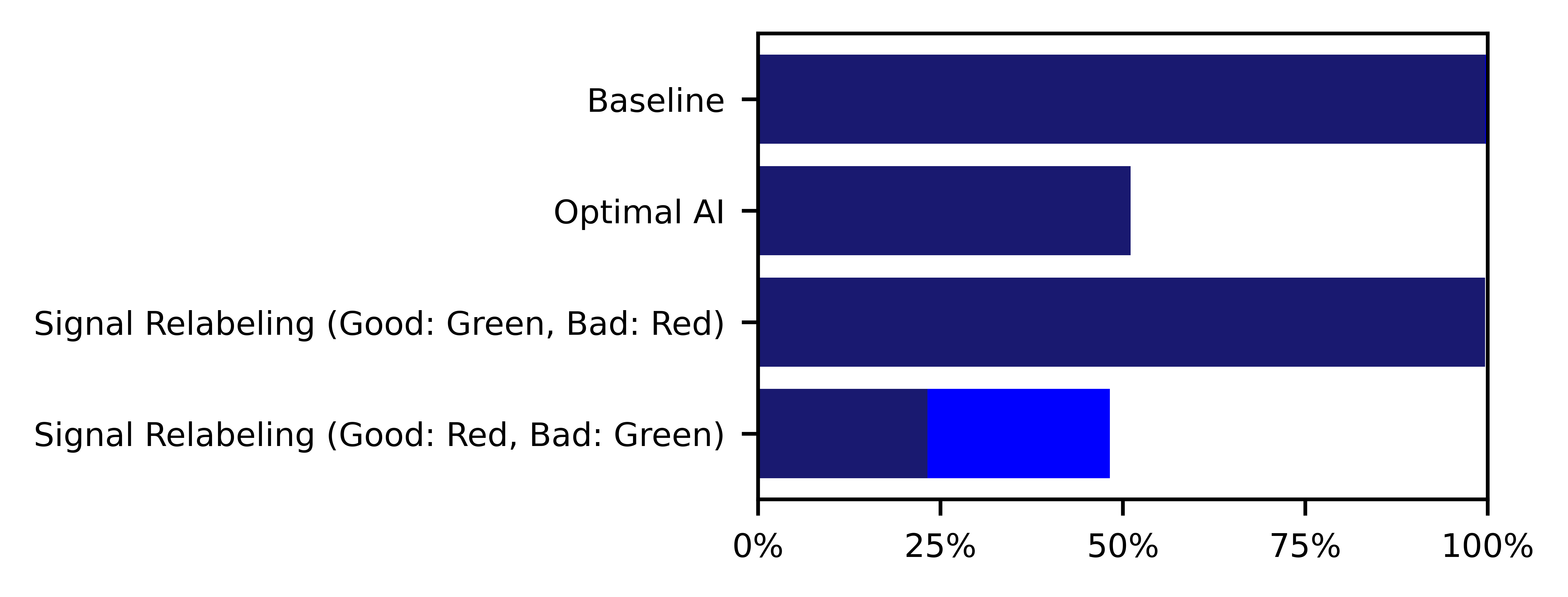}}
\end{figure}
\clearpage

\section*{Tables}

\begin{table}[h!]
    \caption{
        \textbf{Trading behavior in AI and human laboratories}
        \vspace{10pt}\\
        \footnotesize{The table shows the distribution of decisions in the human and AI laboratories. Decisions are averaged across all sessions and trading periods in (a) Treatment I (without event uncertainty), (b) Treatment II (with event uncertainty), and (c) Treatment III (without price updating). \quotation{Human} decisions are taken directly from \citet{cipriani2009herd} for Treatment I and II.\quotation{AI} decisions represent the average decisions across all LLMs. The table also show the results separately for each LLM. \quotation{Rational} behavior represents cases where the informed trader chooses to buy upon receiving a white signal and sell upon receiving a blue signal. \quotation{Partial Rational} behavior represents cases where the  informed trader chooses to buy (sell) upon receiving a white (blue) signal and not trade upon receiving the other signal. \quotation{Cascade Trading} represents cases where the informed trader chooses the same trading action (buy or sell) regardless of the private signal. These decisions are decomposed into \quotation{Optimal Herding}, \quotation{Suboptimal Herding}, \quotation{Contrarian} behavior, and cases where the trade imbalance is zero (\quotation{Undetermined}). While the exact decomposition of human cascade trading decisions is unknown, \citet{cipriani2009herd} show that all types of decisions are present among human traders, as represented by + in the table. \quotation{Cascade No Trading} represents cases where the informed trader chooses not to trade regardless of the private signal. \quotation{Error} represents cases where the informed trader chooses to buy upon receiving a blue signal and sell upon receiving a white signal. The table also reports the frequency of trading periods where herding is optimal. }
    }
    \label{tab:table1_baseline}
    \centering
    \footnotesize{
        \subfloat[Treatment I]{
        \begin{tabular}{lcccccc}
\toprule
 & Human & AI & Claude 3.7 & Claude 3.5 & Llama 3 & Nova Pro \\
\midrule
Rational & 45.70\% & 61.00\% & 70.97\% & 37.11\% & 97.66\% & 38.28\% \\
Partial Rational & 19.60\% & 29.48\% & 11.29\% & 45.31\% & 0.00\% & 61.33\% \\
Cascade Trading & 19.00\% & 9.42\% & 17.74\% & 17.58\% & 2.34\% & 0.00\% \\
\hskip15pt Optimal Herding & 0.00\% & 0.00\% & 0.00\% & 0.00\% & 0.00\% & 0.00\% \\
\hskip15pt Suboptimal Herding & + & 0.00\% & 0.00\% & 0.00\% & 0.00\% & 0.00\% \\
\hskip15pt Contrarian & + & 9.42\% & 17.74\% & 17.58\% & 2.34\% & 0.00\% \\
\hskip15pt Undetermined & + & 0.00\% & 0.00\% & 0.00\% & 0.00\% & 0.00\% \\
Cascade No Trading & 12.30\% & 0.10\% & 0.00\% & 0.00\% & 0.00\% & 0.39\% \\
Error & 3.40\% & 0.00\% & 0.00\% & 0.00\% & 0.00\% & 0.00\% \\
\midrule
Optimal Herding Opportunities & 0.00\% & 0.00\% & 0.00\% & 0.00\% & 0.00\% & 0.00\% \\
\bottomrule
\end{tabular}
}
        \\

        \emph{(Table continues on next page)}
    }
\end{table}
\clearpage

\begin{table}[h!]\ContinuedFloat
    \caption{
        \textbf{Trading behavior in AI and human laboratories} \emph{(continued)}
        \vspace{10pt}\\
    }
    \centering
    \footnotesize{
        \subfloat[Treatment II]{
        \begin{tabular}{lcccccc}
\toprule
 & Human & AI & Claude 3.7 & Claude 3.5 & Llama 3 & Nova Pro \\
\midrule
Rational & 50.90\% & 97.36\% & 100.00\% & 100.00\% & 100.00\% & 89.45\% \\
Partial Rational & 20.10\% & 2.64\% & 0.00\% & 0.00\% & 0.00\% & 10.55\% \\
Cascade Trading & 12.00\% & 0.00\% & 0.00\% & 0.00\% & 0.00\% & 0.00\% \\
\hskip15pt Optimal Herding & + & 0.00\% & 0.00\% & 0.00\% & 0.00\% & 0.00\% \\
\hskip15pt Suboptimal Herding & + & 0.00\% & 0.00\% & 0.00\% & 0.00\% & 0.00\% \\
\hskip15pt Contrarian & + & 0.00\% & 0.00\% & 0.00\% & 0.00\% & 0.00\% \\
\hskip15pt Undetermined & + & 0.00\% & 0.00\% & 0.00\% & 0.00\% & 0.00\% \\
Cascade No Trading & 16.50\% & 0.00\% & 0.00\% & 0.00\% & 0.00\% & 0.00\% \\
Error & 0.05\% & 0.00\% & 0.00\% & 0.00\% & 0.00\% & 0.00\% \\
\midrule
Optimal Herding Opportunities & + & 36.56\% & 30.61\% & 46.88\% & 21.88\% & 46.88\% \\
\bottomrule
\end{tabular}
}\\
        
        \subfloat[Treatment III]{
        \begin{tabular}{lccccc}
\toprule
 & AI & Claude 3.7 & Claude 3.5 & Llama 3 & Nova Pro \\
\midrule
Rational & 99.65\% & 99.38\% & 99.22\% & 100.00\% & 100.00\% \\
Partial Rational & 0.16\% & 0.62\% & 0.00\% & 0.00\% & 0.00\% \\
Cascade Trading & 0.00\% & 0.00\% & 0.00\% & 0.00\% & 0.00\% \\
\hskip15pt Optimal Herding & 0.00\% & 0.00\% & 0.00\% & 0.00\% & 0.00\% \\
\hskip15pt Suboptimal Herding & 0.00\% & 0.00\% & 0.00\% & 0.00\% & 0.00\% \\
\hskip15pt Contrarian & 0.00\% & 0.00\% & 0.00\% & 0.00\% & 0.00\% \\
\hskip15pt Undetermined & 0.00\% & 0.00\% & 0.00\% & 0.00\% & 0.00\% \\
Cascade No Trading & 0.20\% & 0.00\% & 0.78\% & 0.00\% & 0.00\% \\
Error & 0.00\% & 0.00\% & 0.00\% & 0.00\% & 0.00\% \\
\midrule
Optimal Herding Opportunities & 38.36\% & 50.31\% & 34.38\% & 34.38\% & 34.38\% \\
\bottomrule
\end{tabular}
}
    }
\end{table}
\clearpage

\begin{table}[h!]
    \caption{
        \textbf{LLM analysis of AI agent reasoning}
        \vspace{10pt}\\
        \footnotesize{The table shows LLM analysis of reasoning passages using the Claude 3.7 model. The model is prompted to read each passage of reasoning and answer the following five questions. Question 1: Is the trader comparing the price to the expected fundamental value of the asset? (True/False). Question 2: Is the expected value computed using only the signal accuracy and the signal, e.g., 0.7*100+0*0.3=70 or 0.7*0+0.3*100=30? (True/False). Question 3: Does the trader consider the market trend or the trading history in their reasoning? (True/False). Question 4: How does the trader characterize the attractiveness of the investment? Question 5: On a scale from 0-100 (where 100 represents purely emotional and 0 represents purely rational or logical), how much is the investor driven by emotions in their assessment? For \quotation{True/False} questions, the table reports the fraction of \quotation{True} answers.   Panel (a) reports the analysis of reasoning provided by all LLMs and panel (b) reports analysis of reasoning in the Llama 3 model.}
    }
    \label{tab:reasoning_llm}
    \centering
    \footnotesize{
        \subfloat[AI]{\begin{tabular}{lccc}
\toprule
 & Treatment I & Treatment II & Treatment III \\
\midrule
Question 1 & 99.16\% & 99.01\% & 99.67\% \\
Question 2 & 63.07\% & 63.09\% & 63.51\% \\
Question 3 & 17.05\% & 9.50\% & 24.12\% \\
Question 4 &  &  &  \\
\hskip15pt VERY ATTRACTIVE & 11.69\% & 1.88\% & 0.71\% \\
\hskip15pt ATTRACTIVE & 42.83\% & 69.39\% & 68.33\% \\
\hskip15pt REASONABLE & 12.65\% & 6.08\% & 7.33\% \\
\hskip15pt LESS ATTRACTABLE & 10.36\% & 3.65\% & 5.69\% \\
\hskip15pt NO INCENTIVE & 22.17\% & 19.01\% & 17.89\% \\
Question 5 &  &  &  \\
\hskip15pt  Mean & 6.39\% & 4.93\% & 5.62\% \\
\hskip15pt  Bottom decile & 0.00\% & 0.00\% & 0.00\% \\
\hskip15pt  Median & 0.00\% & 0.00\% & 0.00\% \\
\hskip15pt  Top decile & 20.00\% & 15.00\% & 20.00\% \\
\bottomrule
\end{tabular}
}
        \\

        \emph{(Table continues on next page)}
    }
\end{table}
\clearpage

\begin{table}[h!]\ContinuedFloat
    \caption{
        \textbf{LLM analysis of AI agent reasoning} \emph{(continued)}
        \vspace{10pt}\\
    }
    \centering
    \footnotesize{
        \subfloat[Llama 3]{
        \begin{tabular}{lccc}
\toprule
 & Treatment I & Treatment II & Treatment III \\
\midrule
Question 1 & 97.46\% & 100.00\% & 99.61\% \\
Question 2 & 13.87\% & 4.49\% & 4.30\% \\
Question 3 & 40.62\% & 30.66\% & 66.60\% \\
Question 4 &  &  &  \\
\hskip15pt VERY ATTRACTIVE & 1.56\% & 1.56\% & 0.39\% \\
\hskip15pt ATTRACTIVE & 69.92\% & 83.79\% & 74.61\% \\
\hskip15pt REASONABLE & 13.48\% & 4.88\% & 4.49\% \\
\hskip15pt LESS ATTRACTABLE & 10.16\% & 5.66\% & 8.98\% \\
\hskip15pt NO INCENTIVE & 4.69\% & 4.10\% & 11.52\% \\
Question 5 &  &  &  \\
\hskip15pt  Mean & 14.53\% & 12.72\% & 16.65\% \\
\hskip15pt  Bottom decile & 5.00\% & 5.00\% & 10.00\% \\
\hskip15pt  Median & 15.00\% & 10.00\% & 15.00\% \\
\hskip15pt  Top decile & 25.00\% & 20.00\% & 25.00\% \\
\bottomrule
\end{tabular}
}
    }
\end{table}
\clearpage

\begin{table}[h!]
    \caption{
        \textbf{LDA topic analysis of AI agent reasoning}
        \vspace{10pt}\\
        \footnotesize{The table shows the distribution of reasoning passages across LDA topics. The number of topics is fixed to three; using more topics does not result in a larger number of distinct topics. The word clouds associated with the topics are shown in Figure \ref{fig:lda_topics_wordclouds}. Panel (a) reports the analysis of reasoning provided by all LLMs and panel (b) reports analysis of reasoning in the Llama 3 model.}
    }
    \label{tab:reasoning_lda}
    \centering
    \footnotesize{
        \subfloat[AI]{
        \begin{tabular}{lccc}
\toprule
 & Treatment I & Treatment II & Treatment III \\
\midrule
Topic 0 & 44.22\% & 51.93\% & 51.70\% \\
Topic 1 & 26.81\% & 21.27\% & 20.40\% \\
Topic 2 & 28.98\% & 26.80\% & 27.90\% \\
\bottomrule
\end{tabular}
}
        \\
        \subfloat[Llama 3]{
        \begin{tabular}{lccc}
\toprule
 & Treatment I & Treatment II & Treatment III \\
\midrule
Topic 0 & 1.95\% & 0.20\% & 0.59\% \\
Topic 1 & 6.84\% & 9.57\% & 0.59\% \\
Topic 2 & 91.21\% & 90.23\% & 98.83\% \\
\bottomrule
\end{tabular}
}
    }
\end{table}
\clearpage

\begin{table}[h!]
    \caption{
        \textbf{Trading behavior in optimal AI laboratory}
        \vspace{10pt}\\
        \footnotesize{The table shows the distribution of decisions in the optimal AI laboratory in which LLMs are prompted with guidance on optimal decision making. Decisions are averaged across all sessions and trading periods in (a) Treatment I (without event uncertainty), (b) Treatment II (with event uncertainty), and (c) Treatment III (without price updating). \quotation{Optimal AI} decisions represent the average decisions across all LLMs. The table also show the results separately for each LLM. \quotation{Rational} behavior represents cases where the informed trader chooses to buy upon receiving a white signal and sell upon receiving a blue signal. \quotation{Partial Rational} behavior represents cases where the  informed trader chooses to buy (sell) upon receiving a white (blue) signal and not trade upon receiving the other signal. \quotation{Cascade Trading} represents cases where the informed trader chooses the same trading action (buy or sell) regardless of the private signal. These decisions are decomposed into \quotation{Optimal Herding}, \quotation{Suboptimal Herding}, \quotation{Contrarian} behavior, and cases where the trade imbalance is zero (\quotation{Undetermined}). \quotation{Cascade No Trading} represents cases where the informed trader chooses not to trade regardless of the private signal. \quotation{Error} represents cases where the informed trader chooses to buy upon receiving a blue signal and sell upon receiving a white signal. The table also reports the frequency of trading periods where herding is optimal. }
    }
    \label{tab:table1_optimal}
    \centering
    \footnotesize{
        \subfloat[Treatment I]{
        \begin{tabular}{lccccc}
\toprule
 & Optimal AI & Claude 3.7 & Claude 3.5 & Llama 3 & Nova Pro \\
\midrule
Rational & 55.88\% & 48.15\% & 52.34\% & 95.31\% & 27.73\% \\
Partial Rational & 40.60\% & 51.85\% & 38.28\% & 0.00\% & 72.27\% \\
Cascade Trading & 3.52\% & 0.00\% & 9.38\% & 4.69\% & 0.00\% \\
\hskip15pt Optimal Herding & 0.00\% & 0.00\% & 0.00\% & 0.00\% & 0.00\% \\
\hskip15pt Suboptimal Herding & 0.00\% & 0.00\% & 0.00\% & 0.00\% & 0.00\% \\
\hskip15pt Contrarian & 3.52\% & 0.00\% & 9.38\% & 4.69\% & 0.00\% \\
\hskip15pt Undetermined & 0.00\% & 0.00\% & 0.00\% & 0.00\% & 0.00\% \\
Cascade No Trading & 0.00\% & 0.00\% & 0.00\% & 0.00\% & 0.00\% \\
Error & 0.00\% & 0.00\% & 0.00\% & 0.00\% & 0.00\% \\
\midrule
Optimal Herding Opportunities & 0.00\% & 0.00\% & 0.00\% & 0.00\% & 0.00\% \\
\bottomrule
\end{tabular}
}
        \\

        \emph{(Table continues on next page)}
    }
\end{table}
\clearpage

\begin{table}[h!]\ContinuedFloat
    \caption{
        \textbf{Trading behavior in optimal AI laboratory} \emph{(continued)}
        \vspace{10pt}\\
    }
    \centering
    \footnotesize{
        \subfloat[Treatment II]{
        \begin{tabular}{lccccc}
\toprule
 & Optimal AI & Claude 3.7 & Claude 3.5 & Llama 3 & Nova Pro \\
\midrule
Rational & 18.65\% & 37.09\% & 12.50\% & 12.50\% & 12.50\% \\
Partial Rational & 21.88\% & 0.00\% & 0.00\% & 0.00\% & 87.50\% \\
Cascade Trading & 59.48\% & 62.91\% & 87.50\% & 87.50\% & 0.00\% \\
\hskip15pt Optimal Herding & 47.43\% & 39.74\% & 75.00\% & 75.00\% & 0.00\% \\
\hskip15pt Suboptimal Herding & 0.00\% & 0.00\% & 0.00\% & 0.00\% & 0.00\% \\
\hskip15pt Contrarian & 6.60\% & 13.91\% & 6.25\% & 6.25\% & 0.00\% \\
\hskip15pt Undetermined & 5.44\% & 9.27\% & 6.25\% & 6.25\% & 0.00\% \\
Cascade No Trading & 0.00\% & 0.00\% & 0.00\% & 0.00\% & 0.00\% \\
Error & 0.00\% & 0.00\% & 0.00\% & 0.00\% & 0.00\% \\
\midrule
Optimal Herding Opportunities & 81.52\% & 63.58\% & 87.50\% & 87.50\% & 87.50\% \\
\bottomrule
\end{tabular}
}\\
        
        \subfloat[Treatment III]{
        \begin{tabular}{lccccc}
\toprule
 & Optimal AI & Claude 3.7 & Claude 3.5 & Llama 3 & Nova Pro \\
\midrule
Rational & 51.05\% & 45.60\% & 31.25\% & 35.55\% & 91.80\% \\
Partial Rational & 0.00\% & 0.00\% & 0.00\% & 0.00\% & 0.00\% \\
Cascade Trading & 48.95\% & 54.40\% & 68.75\% & 64.45\% & 8.20\% \\
\hskip15pt Optimal Herding & 44.36\% & 53.60\% & 59.38\% & 56.25\% & 8.20\% \\
\hskip15pt Suboptimal Herding & 4.01\% & 0.80\% & 7.03\% & 8.20\% & 0.00\% \\
\hskip15pt Contrarian & 0.00\% & 0.00\% & 0.00\% & 0.00\% & 0.00\% \\
\hskip15pt Undetermined & 0.59\% & 0.00\% & 2.34\% & 0.00\% & 0.00\% \\
Cascade No Trading & 0.00\% & 0.00\% & 0.00\% & 0.00\% & 0.00\% \\
Error & 0.00\% & 0.00\% & 0.00\% & 0.00\% & 0.00\% \\
\midrule
Optimal Herding Opportunities & 50.90\% & 53.60\% & 59.38\% & 56.25\% & 34.38\% \\
\bottomrule
\end{tabular}
}
    }
\end{table}
\clearpage

\begin{table}[h!]
    \caption{
        \textbf{Expected Payoffs}
        \vspace{10pt}\\
        \footnotesize{The table shows descriptive statistics of expected payoffs in the AI and Optimal AI laboratories. In the Optimal AI laboratory, LLMs are prompted with guidance on optimal decision making. Expected payoffs are computed as $\E(v\vert s_t,h_t)-p_t$ if the agent decides to buy the asset, $p_t-\E(v\vert s_t,h_t)$ if the agent decides to sell the asset, and zero otherwise. Expected payoffs are averaged across all sessions, trading periods, and LLMs before computing the statistics.
}
    }
    \label{tab:payoffs}
    \centering
    \footnotesize{
        \begin{tabular}{lcccccccc}
\toprule
&\multicolumn{2}{c}{Treatment I} & &\multicolumn{2}{c}{Treatment II} & &\multicolumn{2}{c}{Treatment III} \\
\cmidrule{2-3}\cmidrule{5-6}\cmidrule{8-9}
 & AI & Optimal AI &  & AI & Optimal AI &  & AI & Optimal AI \\
\midrule
Mean & 2.57 & 2.72 && 3.80 & 14.95 && 5.07 & 7.79 \\
Median & 2.74 & 2.74 && 6.67 & 19.53 && 6.67 & 11.49 \\
Min & -6.67 & -6.67 && -11.44 & -28.28 && -16.19 & -16.19 \\
Max & 6.67 & 6.67 && 11.55 & 28.35 && 16.46 & 16.63 \\
Std  Dev & 3.90 & 3.57 && 6.47 & 14.20 && 8.83 & 7.87 \\
\bottomrule
\end{tabular}

    }
\end{table}
\clearpage

\begin{table}[h!]
    \caption{
        \textbf{Trading behavior in AI laboratory with different types of signals}
        \vspace{10pt}\\
        \footnotesize{The table shows the distribution of decisions in the AI laboratory when varying the color used to code the private information signals. Respectively the white and blue signals are replaced by (i) green and red, and (ii) red and green. Decisions are averaged across all sessions, trading periods, and LLMs in (a) Treatment I (without event uncertainty), (b) Treatment II (with event uncertainty), and (c) Treatment III (without price updating). \quotation{Rational} behavior represents cases where the informed trader chooses to buy upon receiving a white signal and sell upon receiving a blue signal. \quotation{Partial Rational} behavior represents cases where the  informed trader chooses to buy (sell) upon receiving a white (blue) signal and not trade upon receiving the other signal. \quotation{Cascade Trading} represents cases where the informed trader chooses the same trading action (buy or sell) regardless of the private signal. These decisions are decomposed into \quotation{Optimal Herding}, \quotation{Suboptimal Herding}, \quotation{Contrarian} behavior, and cases where the trade imbalance is zero (\quotation{Undetermined}). \quotation{Cascade No Trading} represents cases where the informed trader chooses not to trade regardless of the private signal. \quotation{Error} represents cases where the informed trader chooses to buy upon receiving a blue signal and sell upon receiving a white signal. The table also reports the frequency of trading periods where herding is optimal. }
    }
    \label{tab:table1_signals}
    \centering
    \footnotesize{
        \subfloat[Treatment I]{
        \begin{tabular}{lcc}
\toprule
 & Good: Green, Bad: Red & Good: Red, Bad: Green \\
\midrule
Rational & 54.83\% & 20.23\% \\
Partial Rational & 37.55\% & 42.76\% \\
Cascade Trading & 7.62\% & 6.64\% \\
\hskip15pt Optimal Herding & 0.00\% & 0.00\% \\
\hskip15pt Suboptimal Herding & 0.00\% & 3.52\% \\
\hskip15pt Contrarian & 7.62\% & 0.00\% \\
\hskip15pt Undetermined & 0.00\% & 3.12\% \\
Cascade No Trading & 0.00\% & 5.18\% \\
Error & 0.00\% & 25.20\% \\
\midrule
Optimal Herding Opportunities & 0.00\% & 0.00\% \\
\bottomrule
\end{tabular}
}
        \\
        \emph{(Table continues on next page)}
    }
\end{table}
\clearpage

\begin{table}[h!]\ContinuedFloat
    \caption{
        \textbf{Trading behavior in AI laboratory with different types of signals} \emph{(continued)}
        \vspace{10pt}\\
    }
    \centering
    \footnotesize{
        \subfloat[Treatment II]{
        \begin{tabular}{lcc}
\toprule
 & Good: Green, Bad: Red & Good: Red, Bad: Green \\
\midrule
Rational & 98.54\% & 50.78\% \\
Partial Rational & 1.46\% & 11.72\% \\
Cascade Trading & 0.00\% & 12.50\% \\
\hskip15pt Optimal Herding & 0.00\% & 7.32\% \\
\hskip15pt Suboptimal Herding & 0.00\% & 1.56\% \\
\hskip15pt Contrarian & 0.00\% & 0.00\% \\
\hskip15pt Undetermined & 0.00\% & 3.61\% \\
Cascade No Trading & 0.00\% & 0.00\% \\
Error & 0.00\% & 25.00\% \\
\midrule
Optimal Herding Opportunities & 52.94\% & 42.93\% \\
\bottomrule
\end{tabular}
}\\
        
        \subfloat[Treatment III]{
        \begin{tabular}{lcc}
\toprule
 & Good: Green, Bad: Red & Good: Red, Bad: Green \\
\midrule
Rational & 99.64\% & 23.24\% \\
Partial Rational & 0.00\% & 25.00\% \\
Cascade Trading & 0.36\% & 25.18\% \\
\hskip15pt Optimal Herding & 0.36\% & 18.82\% \\
\hskip15pt Suboptimal Herding & 0.00\% & 3.14\% \\
\hskip15pt Contrarian & 0.00\% & 0.18\% \\
\hskip15pt Undetermined & 0.00\% & 3.04\% \\
Cascade No Trading & 0.00\% & 0.00\% \\
Error & 0.00\% & 26.58\% \\
\midrule
Optimal Herding Opportunities & 56.61\% & 56.75\% \\
\bottomrule
\end{tabular}

        }
    }
\end{table}
\clearpage

\clearpage
\section*{Prompts}
\setcounter{figure}{0}
\renewcommand{\figurename}{Prompt}

\begin{figure}[h!]
    \centering
    \caption{\textbf{System prompt}
    \vspace{5pt}\\
    \footnotesize{This prompt describes the instructions of the experiment, which is given to the LLMs through their system prompt. }
    \vspace{5pt}
    }
    \label{prmpt:system} 

    \begin{tcolorbox}[colback=gray!10, colframe=gray!10, width=\textwidth]
    \scriptsize
    \begin{Verbatim}[commandchars=\\\{\},breaklines=true, breakanywhere=true]
    You are participating in an experiment at the Experimental Laboratory of the ELSE Centre at the Department of Economics at UCL. The instructions given for the laborary experiment are as follows:
    
    There are a total of 8 participants in this experiment. Everyone is receiving the same instructions.
    
    In the experiment, you can exchange one unit of an asset with a computerized market maker. You and the other participants will make trading decisions through 8 sequential rounds. In each round, only one participant will be selected to trade. Each participant can only trade once.
    
    In each round, the market maker sets the price of the asset as the expected value of the fundamental value of the asset, conditional on the history of the trades from the previous rounds.
    
    [\textcolor{blue}{if treatment==2: \{The market maker will update the price as if, with high probability, it were trading not with informed traders, but with noise traders.}\textcolor{blue}{\}}]
    
    The fundamental value of the asset is a discrete random variable that can take values 0 or 100, each with a 50% probability. You do not know the fundamental value of the asset, but you may receive a signal (white or blue) on the value. If the asset value is 100, you receive a white signal with 70% probability and a blue signal with 30% probability. If the value is 0, you receive a white signal with 30% probability and a blue signal with 70% probability.
      
    \end{Verbatim}
    \end{tcolorbox}
    \textit{Prompt continues on next page.}
\end{figure}
\clearpage

\begin{figure}[h!]\ContinuedFloat
    \centering
    \caption{\textbf{System prompt} \textit{(continued)}
    \vspace{5pt}
    }

    \begin{tcolorbox}[colback=gray!10, colframe=gray!10, width=\textwidth]
    \scriptsize
    \begin{Verbatim}[commandchars=\\\{\},breaklines=true, breakanywhere=true]
    You will be making decisions on whether to buy or sell one unit of the asset at a given price, or not to trade given respectively a white and a blue signal. The realized signal will only be revealed to you if you are selected to trade. After each round, the computer will randomly select a participant whose trade gets executed. That participant receives the realized signal. The remaining participants then observe the executed trading decision (buy, sell, or no trade), but do not receive the realized signal. They also do not observe the identity of the selected participant. The procedure continues for 8 rounds until all participants have acted once. All participants (including those whose decision has already been executed) observe the trading decisions in each period and the corresponding price movement. 
        
    After 8 rounds, the asset value is revealed, and each participant receives a payoff computed based on the trading decision and price in the round in which the participant was selected and the asset value v.

    Payoffs are computed in a fictitious experimental currency called lira. If the participant sold the asset at price p, the payoff is p-v lire. If the participant bought the asset at price p, the payoff is v-p lire. If the participant decided not to trade, the payoff is zero lire. At the end of the experiment, the payoffs are added up and converted into British pounds at the rate of 3 lire per pound. In addition, you are paid 70 pounds for participating in the experiment, regardless of your payoff.        
    \end{Verbatim}
    \end{tcolorbox}
\end{figure}
\clearpage

\begin{figure}[h!]
    \centering
    \caption{\textbf{User prompt in AI laboratory}
    \vspace{5pt}\\
    \footnotesize{This prompt describes the instructions given to the LLMs in each trading period to each agent $j$. The HISTORY input consists of the executed trades of selected traders along with the history of actions and reasoning for agent $j$ in all previous periods. In addition to this user prompt, the LLMs have available the instructions through the system prompt, see Prompt \ref{prmpt:system}.}
    \vspace{5pt}
    }
    \label{prmpt:user} 

    \begin{tcolorbox}[colback=gray!10, colframe=gray!10, width=\textwidth]
    \scriptsize
    \begin{Verbatim}[commandchars=\\\{\},breaklines=true, breakanywhere=true]
    This is round [\textcolor{blue}{TRADING PERIOD (t)}].

    [\textcolor{blue}{HISTORY}]
    
    If you receive a white signal, will you buy, sell, or not trade at a price of [\textcolor{blue}{PRICE}]? 
    If you receive a blue signal, will you buy, sell, or not trade at a price of [\textcolor{blue}{PRICE}]? 
    
    Please make sure that you provide your response in the following format:
    \{
        "actionWhite": "BUY/SELL/NO TRADE at the price of [\textcolor{blue}{PRICE}] conditional on observing a white signal",
        "actionBlue": "BUY/SELL/NO TRADE at the price of [\textcolor{blue}{PRICE}] conditional on observing a blue signal",
        "reasoningWhite": "Brief explanation of your decision conditional on observing a white signal (1-2 sentences) ",
        "reasoningBlue": "Brief explanation of your decision conditional on observing a blue signal (1-2 sentences)"
    \}
    \end{Verbatim}
    \end{tcolorbox}
\end{figure}
\clearpage

\begin{figure}[h!]
    \centering
    \caption{\textbf{User prompt in optimal AI laboratory}
    \vspace{5pt}\\
    \footnotesize{This prompt describes the instructions given to the LLMs in each trading period to each agent $j$ in the optimal AI laboratory. The HISTORY input consists of the executed trades of selected traders along with the history of actions and reasoning for agent $j$ in all previous periods. In addition to this user prompt, the LLMs have available the instructions through the system prompt, see Prompt \ref{prmpt:system}.}
    \vspace{5pt}
    }
    \label{prmpt:user_optimal} 

    \begin{tcolorbox}[colback=gray!10, colframe=gray!10, width=\textwidth]
    \scriptsize
    \begin{Verbatim}[commandchars=\\\{\},breaklines=true, breakanywhere=true]
    This is round [\textcolor{blue}{TRADING PERIOD (t)}].

    \textcolor{blue}{if TRADING PERIOD (t)==1:}
        Note that given current conditions, it is optimal to buy given a white signal and sell given a blue signal.
    
    \textcolor{blue}{else:}
        [\textcolor{blue}{HISTORY}]
        
        \textcolor{blue}{if expected_value_trader_white > price and expected_value_trader_blue > price:}
            Note that given current conditions, it is optimal to follow the herd and buy regardless of the signal.

        \textcolor{blue}{if expected_value_trader_white < price and expected_value_trader_blue < price::}  
            Note that given current conditions, it is optimal to follow the herd and sell regardless of the signal.
     
    If you receive a white signal, will you buy, sell, or not trade at a price of [\textcolor{blue}{PRICE}]? 
    If you receive a blue signal, will you buy, sell, or not trade at a price of [\textcolor{blue}{PRICE}]? 
    
    Please make sure that you provide your response in the following format:
    \{
        "actionWhite": "BUY/SELL/NO TRADE at the price of [\textcolor{blue}{PRICE}] conditional on observing a white signal",
        "actionBlue": "BUY/SELL/NO TRADE at the price of [\textcolor{blue}{PRICE}] conditional on observing a blue signal",
        "reasoningWhite": "Brief explanation of your decision conditional on observing a white signal (1-2 sentences) ",
        "reasoningBlue": "Brief explanation of your decision conditional on observing a blue signal (1-2 sentences)"
    \}
    \end{Verbatim}
    \end{tcolorbox}
\end{figure}
\clearpage

\begin{figure}[h!]
    \centering
    \caption{\textbf{System prompt personal characteristics add-on}
    \vspace{5pt}\\
    \footnotesize{This prompt describes an add-on to the system prompt that provides characteristics of the AI agent. The characteristics are drawn randomly from the unconditional distributions of human participant characteristics reported in \citet{cipriani2009herd} restricted according to a set of heuristics to ensure realistic personas.}
    \vspace{5pt}
    }
    \label{prmpt:system_characteristics_addon} 

    \begin{tcolorbox}[colback=gray!10, colframe=gray!10, width=\textwidth]
    \scriptsize
    \begin{Verbatim}[commandchars=\\\{\},breaklines=true, breakanywhere=true]
    You are a [\textcolor{blue}{AGE}]-year old [\textcolor{blue}{GENDER}]. You work as a [\textcolor{blue}{OCCUPATION}] and you have [\textcolor{blue}{TENURE}] years of tenure. You have a [\textcolor{blue}{EDUCATION LEVEL}] degree in [\textcolor{blue}{EDUCATION FIELD}]. Respond in way that is consistent with the knowledge and expected behavior of a person with these characteristics.
    \end{Verbatim}
    \end{tcolorbox}
\end{figure}
\clearpage


\clearpage
\begin{appendices}

\section{Robustness to temperature}\label{app:temperature}
\begin{table}[h!]
    \caption{
        \textbf{Trading behavior in AI laboratory with different model temperature settings}
        \vspace{10pt}\\
        \footnotesize{The table shows the distribution of decisions in the AI laboratory with varying temperatures for the Claude 3.5, Llama 3, and Nova Pro models (the temperature is fixed at one in the Claude 3.7 reasoning model). Decisions are averaged across all sessions, trading periods, and LLMs in (a) Treatment I (without event uncertainty), (b) Treatment II (with event uncertainty), and (c) Treatment III (without price updating). \quotation{Rational} behavior represents cases where the informed trader chooses to buy upon receiving a white signal and sell upon receiving a blue signal. \quotation{Partial Rational} behavior represents cases where the  informed trader chooses to buy (sell) upon receiving a white (blue) signal and not trade upon receiving the other signal. \quotation{Cascade Trading} represents cases where the informed trader chooses the same trading action (buy or sell) regardless of the private signal. These decisions are decomposed into \quotation{Optimal Herding}, \quotation{Suboptimal Herding}, \quotation{Contrarian} behavior, and cases where the trade imbalance is zero (\quotation{Undetermined}). \quotation{Cascade No Trading} represents cases where the informed trader chooses not to trade regardless of the private signal. \quotation{Error} represents cases where the informed trader chooses to buy upon receiving a blue signal and sell upon receiving a white signal. The table also reports the frequency of trading periods where herding is optimal.}
    }
    \label{tab:table1_temperatures}
    \centering
    \footnotesize{
        \subfloat[Treatment I]{
        \begin{tabular}{lccc}
\toprule
 & T=0.0 & T=0.7 (baseline) & T=1.0 \\
\midrule
Rational & 58.95\% & 61.00\% & 66.28\% \\
Partial Rational & 32.22\% & 29.48\% & 27.14\% \\
Cascade Trading & 8.83\% & 9.42\% & 6.58\% \\
\hskip15pt Optimal Herding & 0.00\% & 0.00\% & 0.00\% \\
\hskip15pt Suboptimal Herding & 0.00\% & 0.00\% & 0.00\% \\
\hskip15pt Contrarian & 8.83\% & 9.42\% & 6.58\% \\
\hskip15pt Undetermined & 0.00\% & 0.00\% & 0.00\% \\
Cascade No Trading & 0.00\% & 0.10\% & 0.00\% \\
Error & 0.00\% & 0.00\% & 0.00\% \\
\midrule
Optimal Herding Opportunities & 0.00\% & 0.00\% & 0.00\% \\
\bottomrule
\end{tabular}
}
        \\
        \emph{(Table continues on next page)}
    }
\end{table}
\clearpage

\begin{table}[h!]\ContinuedFloat
    \caption{
        \textbf{Trading behavior in AI laboratory with different model temperature settings} \emph{(continued)}
        \vspace{10pt}\\
    }
    \centering
    \footnotesize{
        \subfloat[Treatment II]{
        \begin{tabular}{lccc}
\toprule
 & T=0.0 & T=0.7 (baseline) & T=1.0 \\
\midrule
Rational & 97.27\% & 97.36\% & 88.48\% \\
Partial Rational & 2.73\% & 2.64\% & 11.52\% \\
Cascade Trading & 0.00\% & 0.00\% & 0.00\% \\
\hskip15pt Optimal Herding & 0.00\% & 0.00\% & 0.00\% \\
\hskip15pt Suboptimal Herding & 0.00\% & 0.00\% & 0.00\% \\
\hskip15pt Contrarian & 0.00\% & 0.00\% & 0.00\% \\
\hskip15pt Undetermined & 0.00\% & 0.00\% & 0.00\% \\
Cascade No Trading & 0.00\% & 0.00\% & 0.00\% \\
Error & 0.00\% & 0.00\% & 0.00\% \\
\midrule
Optimal Herding Opportunities & 41.25\% & 36.56\% & 45.15\% \\
\bottomrule
\end{tabular}
}\\
        
        \subfloat[Treatment III]{
        \begin{tabular}{lccc}
\toprule
 & T=0.0 & T=0.7 (baseline) & T=1.0 \\
\midrule
Rational & 99.84\% & 99.65\% & 99.84\% \\
Partial Rational & 0.16\% & 0.16\% & 0.16\% \\
Cascade Trading & 0.00\% & 0.00\% & 0.00\% \\
\hskip15pt Optimal Herding & 0.00\% & 0.00\% & 0.00\% \\
\hskip15pt Suboptimal Herding & 0.00\% & 0.00\% & 0.00\% \\
\hskip15pt Contrarian & 0.00\% & 0.00\% & 0.00\% \\
\hskip15pt Undetermined & 0.00\% & 0.00\% & 0.00\% \\
Cascade No Trading & 0.00\% & 0.20\% & 0.00\% \\
Error & 0.00\% & 0.00\% & 0.00\% \\
\midrule
Optimal Herding Opportunities & 55.55\% & 38.36\% & 50.86\% \\
\bottomrule
\end{tabular}

        }
    }
\end{table}
\clearpage

\section{Additional results}\label{app:additional_results}

\begin{table}[h!]
    \caption{
        \textbf{LLM color connotations}
        \vspace{10pt}\\
        \footnotesize{The table shows how each LLM \quotation{think} about the colors (a) white, (b) blue, (c) green, and (d) red in the context of financial analysis. The prompt used to generate these responses is: \quotation{You are a financial analyst. What do you think a signal color-coded [\textcolor{blue}{color}] says about market conditions? Please respond in a 1-2 sentences.}    
        }
    }
    \label{tab:color_connotation}
    \centering
    \footnotesize{
        \subfloat[White]{
        \begin{tabular}{lp{13.5cm}}
\toprule
Model & Connotation \\
\midrule
Claude 3.7 & In financial analysis, a white signal typically indicates neutral market conditions, suggesting neither bullish nor bearish sentiment predominates. This color code often represents a period of stability or consolidation where prices are moving sideways, prompting analysts to watch for developing patterns before making strong directional recommendations.\\
Claude 3.5 & In financial analysis, a white signal typically indicates neutral or unclear market conditions. It suggests that there is no strong trend or definitive direction in the market, and investors may need to exercise caution and gather more information before making decisions.\\
Llama 3 & 
As a financial analyst, I would interpret a signal color-coded white as a neutral or calm market condition, indicating a lack of strong trends or volatility, and suggesting that the market is consolidating or trading within a narrow range. This could be a sign that investors are awaiting new information or catalysts to drive market direction.\\
Nova Pro & A white signal typically indicates neutral or stable market conditions with no significant bullish or bearish trends.\\
\bottomrule
\end{tabular}}
        \bigskip
    
        \emph{(Table continues on next page)}
    }
\end{table}
\clearpage

\begin{table}[h!]\ContinuedFloat
    \caption{
        \textbf{LLM color connotations} \emph{(continued)}
        \vspace{10pt}\\
    }
    \centering
    \footnotesize{
        \subfloat[Blue]{
        \begin{tabular}{lp{13.5cm}}
\toprule
Model & Connotation \\
\midrule
Claude 3.7 & In financial analysis, a blue signal typically indicates stable or neutral market conditions, suggesting a period of consolidation or sideways movement rather than significant bullish or bearish trends. This often represents a time when investors might want to maintain current positions while waiting for clearer directional indicators to emerge.\\
Claude 3.5 & In financial analysis, a blue signal typically indicates stable or neutral market conditions. It suggests that the market is neither particularly bullish nor bearish, and investors might expect a period of consolidation or steady performance.\\
Llama 3 & 
A signal color-coded blue typically indicates a bullish or positive market signal, suggesting that market conditions are favorable and prices are likely to rise. This could be due to strong economic indicators, increasing demand, or a decrease in volatility, prompting investors to take a more optimistic view of the market.\\
Nova Pro & A blue signal typically indicates stability or a neutral market condition, suggesting no significant upward or downward trends.\\
\bottomrule
\end{tabular}}
        \\
        \bigskip
        
        \subfloat[Green]{
        \begin{tabular}{lp{13.5cm}}
\toprule
Model & Connotation \\
\midrule
Claude 3.7 & A green signal typically indicates positive market conditions, suggesting upward trends, growth opportunities, or favorable economic indicators. Investors often interpret green signals as a time to consider buying or maintaining positions in the market.\\
Claude 3.5 & A green signal in financial markets typically indicates positive or bullish conditions. It suggests that prices are rising, investor sentiment is optimistic, and it may be a favorable time to buy or hold assets.\\
Llama 3 & 
A signal color-coded green typically indicates a bullish market sentiment, suggesting that market conditions are favorable and trending upward, with prices likely to continue rising. This can be a sign of strong economic growth, low volatility, and increased investor confidence, making it a good time to consider buying or holding onto stocks.\\
Nova Pro & A green signal typically indicates a bullish market, suggesting positive momentum and rising asset prices.\\
\bottomrule
\end{tabular}}
    }
\end{table}
\clearpage

\begin{table}[h!]\ContinuedFloat
    \caption{
        \textbf{LLM color connotations} \emph{(continued)}
        \vspace{10pt}\\
    }
    \centering
    \footnotesize{
        \subfloat[Red]{
        \begin{tabular}{lp{13.5cm}}
\toprule
Model & Connotation \\
\midrule
Claude 3.7 & In financial analysis, a red signal typically indicates negative market conditions such as declining prices, bearish sentiment, or heightened risk factors. This warning sign suggests investors should exercise caution and potentially reassess their positions to mitigate downside exposure.\\
Claude 3.5 & A signal color-coded red in financial markets typically indicates negative or bearish conditions. It often suggests a downturn, loss, or potential risk in the market, prompting investors and analysts to exercise caution or consider defensive strategies.\\
Llama 3 & 
A signal color-coded red typically indicates a bearish or cautionary signal, suggesting that market conditions are unfavorable or volatile, and investors should exercise caution or consider taking defensive measures to mitigate potential losses. It may also signal a potential downturn or correction in the market, prompting analysts and investors to reassess their positions and strategies.\\
Nova Pro &A red signal typically indicates bearish market conditions, suggesting a downward trend in asset prices.\\
\bottomrule
\end{tabular}}
    }
\end{table}
\clearpage

\begin{landscape}
\begin{table}[h!]
    \caption{
        \textbf{Trading behavior in AI laboratory with different types of signals by LLM} 
        \vspace{10pt}\\
        \footnotesize{
        The table shows the distribution of decisions in the AI laboratory for each LLM separately when varying the color used to code the private information signals. Respectively the white and blue signals are replaced by (i) green and red, and (ii) red and green. Decisions are averaged across all sessions and trading periods in (a) Treatment I (without event uncertainty),  (b) Treatment II (with event uncertainty), and (c) Treatment III (without price updating). \quotation{Rational} behavior represents cases where the informed trader chooses to buy upon receiving a white signal and sell upon receiving a blue signal. \quotation{Partial Rational} behavior represents cases where the  informed trader chooses to buy (sell) upon receiving a white (blue) signal and not trade upon receiving the other signal. \quotation{Cascade Trading} represents cases where the informed trader chooses the same trading action (buy or sell) regardless of the private signal. These decisions are decomposed into \quotation{Optimal Herding}, \quotation{Suboptimal Herding}, \quotation{Contrarian} behavior, and cases where the trade imbalance is zero (\quotation{Undetermined}). \quotation{Cascade No Trading} represents cases where the informed trader chooses not to trade regardless of the private signal. \quotation{Error} represents cases where the informed trader chooses to buy upon receiving a blue signal and sell upon receiving a white signal. The table also reports the frequency of trading periods where herding is optimal.}
    }
    \label{tab:table1_signals_individual_LLM}
    \centering
    \footnotesize{
        \subfloat[Treatment I]{
        \begin{tabular}{lccccccccc}
\toprule
&\multicolumn{4}{c}{Good: Green, Bad: Red} & &\multicolumn{4}{c}{Good: Red, Bad: Green} \\
\cmidrule{2-5}\cmidrule{7-10}
 & Claude 3.7 & Claude 3.5 & Llama 3 & Nova Pro &  & Claude 3.7 & Claude 3.5 & Llama 3 & Nova Pro \\
\midrule
Rational & 45.86\% & 35.55\% & 100.00\% & 37.89\% && 63.33\% & 0.00\% & 16.80\% & 0.78\% \\
Partial Rational & 34.59\% & 53.52\% & 0.00\% & 62.11\% && 36.67\% & 0.00\% & 46.88\% & 87.50\% \\
Cascade Trading & 19.55\% & 10.94\% & 0.00\% & 0.00\% && 0.00\% & 0.00\% & 26.56\% & 0.00\% \\
\hskip15pt Optimal Herding & 0.00\% & 0.00\% & 0.00\% & 0.00\% && 0.00\% & 0.00\% & 0.00\% & 0.00\% \\
\hskip15pt Suboptimal Herding & 0.00\% & 0.00\% & 0.00\% & 0.00\% && 0.00\% & 0.00\% & 14.06\% & 0.00\% \\
\hskip15pt Contrarian & 19.55\% & 10.94\% & 0.00\% & 0.00\% && 0.00\% & 0.00\% & 0.00\% & 0.00\% \\
\hskip15pt Undetermined & 0.00\% & 0.00\% & 0.00\% & 0.00\% && 0.00\% & 0.00\% & 12.50\% & 0.00\% \\
Cascade No Trading & 0.00\% & 0.00\% & 0.00\% & 0.00\% && 0.00\% & 0.00\% & 8.98\% & 11.72\% \\
Error & 0.00\% & 0.00\% & 0.00\% & 0.00\% && 0.00\% & 100.00\% & 0.78\% & 0.00\% \\
\midrule
Optimal Herding Opportunities & 0.00\% & 0.00\% & 0.00\% & 0.00\% && 0.00\% & 0.00\% & 0.00\% & 0.00\% \\
\bottomrule
\end{tabular}
}
        \\
        \emph{(Table continues on next page)}
    }
\end{table}
\end{landscape}
\clearpage

\begin{landscape}
\begin{table}[h!]\ContinuedFloat
    \caption{
        \textbf{Trading behavior in AI laboratory with different types of signals by LLM} \emph{(continued)}
        \vspace{10pt}\\
    }
    \centering
    \footnotesize{
        \subfloat[Treatment II]{
        \begin{tabular}{lccccccccc}
\toprule
&\multicolumn{4}{c}{Good: Green, Bad: Red} & &\multicolumn{4}{c}{Good: Red, Bad: Green} \\
\cmidrule{2-5}\cmidrule{7-10}
 & Claude 3.7 & Claude 3.5 & Llama 3 & Nova Pro &  & Claude 3.7 & Claude 3.5 & Llama 3 & Nova Pro \\
\midrule
Rational & 100.00\% & 100.00\% & 100.00\% & 94.14\% && 100.00\% & 0.00\% & 3.12\% & 100.00\% \\
Partial Rational & 0.00\% & 0.00\% & 0.00\% & 5.86\% && 0.00\% & 0.00\% & 46.88\% & 0.00\% \\
Cascade Trading & 0.00\% & 0.00\% & 0.00\% & 0.00\% && 0.00\% & 0.00\% & 50.00\% & 0.00\% \\
\hskip15pt Optimal Herding & 0.00\% & 0.00\% & 0.00\% & 0.00\% && 0.00\% & 0.00\% & 29.30\% & 0.00\% \\
\hskip15pt Suboptimal Herding & 0.00\% & 0.00\% & 0.00\% & 0.00\% && 0.00\% & 0.00\% & 6.25\% & 0.00\% \\
\hskip15pt Contrarian & 0.00\% & 0.00\% & 0.00\% & 0.00\% && 0.00\% & 0.00\% & 0.00\% & 0.00\% \\
\hskip15pt Undetermined & 0.00\% & 0.00\% & 0.00\% & 0.00\% && 0.00\% & 0.00\% & 14.45\% & 0.00\% \\
Cascade No Trading & 0.00\% & 0.00\% & 0.00\% & 0.00\% && 0.00\% & 0.00\% & 0.00\% & 0.00\% \\
Error & 0.00\% & 0.00\% & 0.00\% & 0.00\% && 0.00\% & 100.00\% & 0.00\% & 0.00\% \\
\midrule
Optimal Herding Opportunities & 49.28\% & 46.88\% & 46.88\% & 68.75\% && 31.08\% & 40.62\% & 59.38\% & 40.62\% \\
\bottomrule
\end{tabular}
}
        \\
        \emph{(Table continues on next page)}
    }
\end{table}
\end{landscape}
\clearpage

\begin{landscape}
\begin{table}[h!]\ContinuedFloat
    \caption{
        \textbf{Trading behavior in AI laboratory with different types of signals by LLM} \emph{(continued)}
        \vspace{10pt}\\
    }
    \centering
    \footnotesize{
        \subfloat[Treatment III]{
        \begin{tabular}{lccccccccc}
\toprule
&\multicolumn{4}{c}{Good: Green, Bad: Red} & &\multicolumn{4}{c}{Good: Red, Bad: Green} \\
\cmidrule{2-5}\cmidrule{7-10}
 & Claude 3.7 & Claude 3.5 & Llama 3 & Nova Pro &  & Claude 3.7 & Claude 3.5 & Llama 3 & Nova Pro \\
\midrule
Rational & 98.57\% & 100.00\% & 100.00\% & 100.00\% && 92.96\% & 0.00\% & 0.00\% & 0.00\% \\
Partial Rational & 0.00\% & 0.00\% & 0.00\% & 0.00\% && 0.00\% & 0.00\% & 0.00\% & 100.00\% \\
Cascade Trading & 1.43\% & 0.00\% & 0.00\% & 0.00\% && 0.70\% & 0.00\% & 100.00\% & 0.00\% \\
\hskip15pt Optimal Herding & 1.43\% & 0.00\% & 0.00\% & 0.00\% && 0.00\% & 0.00\% & 75.29\% & 0.00\% \\
\hskip15pt Suboptimal Herding & 0.00\% & 0.00\% & 0.00\% & 0.00\% && 0.00\% & 0.00\% & 12.55\% & 0.00\% \\
\hskip15pt Contrarian & 0.00\% & 0.00\% & 0.00\% & 0.00\% && 0.70\% & 0.00\% & 0.00\% & 0.00\% \\
\hskip15pt Undetermined & 0.00\% & 0.00\% & 0.00\% & 0.00\% && 0.00\% & 0.00\% & 12.16\% & 0.00\% \\
Cascade No Trading & 0.00\% & 0.00\% & 0.00\% & 0.00\% && 0.00\% & 0.00\% & 0.00\% & 0.00\% \\
Error & 0.00\% & 0.00\% & 0.00\% & 0.00\% && 6.34\% & 100.00\% & 0.00\% & 0.00\% \\
\midrule
Optimal Herding Opportunities & 51.43\% & 53.12\% & 53.12\% & 68.75\% && 48.59\% & 50.00\% & 75.29\% & 53.12\% \\
\bottomrule
\end{tabular}

        }
    }
\end{table}
\end{landscape}
\clearpage

\begin{table}[h!]
    \caption{
        \textbf{Trading behavior of AI agents with personal profiles}
        \vspace{10pt}\\
        \footnotesize{The table shows the distribution of decisions in the AI laboratory when endowing LLMs with personal profiles (\quotation{Human,} \quotation{Professional Trader,} \quotation{Robo-Advisor,} and \quotation{Rational}) or with personal characteristics drawn from the unconditional distributions of human participants from \citet{cipriani2009herd} subject to realistic constraints (\quotation{C\&G Characteristics}). Decisions are averaged across all sessions, trading periods, and LLMs in (a) Treatment I (without event uncertainty), (b) Treatment II (with event uncertainty), and (c) Treatment III (without price updating). \quotation{Rational} behavior represents cases where the informed trader chooses to buy upon receiving a white signal and sell upon receiving a blue signal. \quotation{Partial Rational} behavior represents cases where the  informed trader chooses to buy (sell) upon receiving a white (blue) signal and not trade upon receiving the other signal. \quotation{Cascade Trading} represents cases where the informed trader chooses the same trading action (buy or sell) regardless of the private signal. These decisions are decomposed into \quotation{Optimal Herding}, \quotation{Suboptimal Herding}, \quotation{Contrarian} behavior, and cases where the trade imbalance is zero (\quotation{Undetermined}). \quotation{Cascade No Trading} represents cases where the informed trader chooses not to trade regardless of the private signal. \quotation{Error} represents cases where the informed trader chooses to buy upon receiving a blue signal and sell upon receiving a white signal. The table also reports the frequency of trading periods where herding is optimal. }
    }
    \label{tab:table1_profiles}
    \centering
    \footnotesize{
        \subfloat[Treatment I]{
        \begin{tabular}{lccccc}
\toprule
 & Human & Professional Trader & Robo-Advisor & Rational & C\&G Characteristics \\
\midrule
Rational & 89.68\% & 67.30\% & 54.21\% & 59.22\% & 59.35\% \\
Partial Rational & 7.69\% & 29.51\% & 37.41\% & 30.66\% & 31.30\% \\
Cascade Trading & 2.63\% & 2.32\% & 5.93\% & 7.88\% & 9.35\% \\
\hskip15pt Optimal Herding & 0.00\% & 0.00\% & 0.00\% & 0.00\% & 0.00\% \\
\hskip15pt Suboptimal Herding & 0.00\% & 0.00\% & 0.00\% & 0.00\% & 0.00\% \\
\hskip15pt Contrarian & 2.63\% & 2.32\% & 5.93\% & 7.88\% & 9.35\% \\
\hskip15pt Undetermined & 0.00\% & 0.00\% & 0.00\% & 0.00\% & 0.00\% \\
Cascade No Trading & 0.00\% & 0.88\% & 2.44\% & 2.25\% & 0.00\% \\
Error & 0.00\% & 0.00\% & 0.00\% & 0.00\% & 0.00\% \\
\midrule
Optimal Herding Opportunities & 0.00\% & 0.00\% & 0.00\% & 0.00\% & 0.00\% \\
\bottomrule
\end{tabular}
}
        \\
        \emph{(Table continues on next page)}
    }
\end{table}
\clearpage

\begin{table}[h!]\ContinuedFloat
    \caption{
        \textbf{Trading behavior of AI agents with personal profiles} \emph{(continued)}
        \vspace{10pt}\\
    }
    \centering
    \footnotesize{
        \subfloat[Treatment II]{
        \begin{tabular}{lccccc}
\toprule
 & Human & Professional Trader & Robo-Advisor & Rational & C\&G Characteristics \\
\midrule
Rational & 93.93\% & 96.97\% & 98.24\% & 93.77\% & 97.36\% \\
Partial Rational & 5.97\% & 3.03\% & 0.59\% & 4.98\% & 2.64\% \\
Cascade Trading & 0.10\% & 0.00\% & 0.49\% & 1.15\% & 0.00\% \\
\hskip15pt Optimal Herding & 0.10\% & 0.00\% & 0.39\% & 0.10\% & 0.00\% \\
\hskip15pt Suboptimal Herding & 0.00\% & 0.00\% & 0.00\% & 0.00\% & 0.00\% \\
\hskip15pt Contrarian & 0.00\% & 0.00\% & 0.00\% & 1.06\% & 0.00\% \\
\hskip15pt Undetermined & 0.00\% & 0.00\% & 0.10\% & 0.00\% & 0.00\% \\
Cascade No Trading & 0.00\% & 0.00\% & 0.00\% & 0.00\% & 0.00\% \\
Error & 0.00\% & 0.00\% & 0.68\% & 0.10\% & 0.00\% \\
\midrule
Optimal Herding Opportunities & 46.26\% & 35.68\% & 44.13\% & 49.98\% & 72.00\% \\
\bottomrule
\end{tabular}
}\\
        
        \subfloat[Treatment III]{
        \begin{tabular}{lccccc}
\toprule
 & Human & Professional Trader & Robo-Advisor & Rational & C\&G Characteristics \\
\midrule
Rational & 93.85\% & 100.00\% & 100.00\% & 98.77\% & 99.90\% \\
Partial Rational & 5.66\% & 0.00\% & 0.00\% & 0.77\% & 0.10\% \\
Cascade Trading & 0.00\% & 0.00\% & 0.00\% & 0.46\% & 0.00\% \\
\hskip15pt Optimal Herding & 0.00\% & 0.00\% & 0.00\% & 0.46\% & 0.00\% \\
\hskip15pt Suboptimal Herding & 0.00\% & 0.00\% & 0.00\% & 0.00\% & 0.00\% \\
\hskip15pt Contrarian & 0.00\% & 0.00\% & 0.00\% & 0.00\% & 0.00\% \\
\hskip15pt Undetermined & 0.00\% & 0.00\% & 0.00\% & 0.00\% & 0.00\% \\
Cascade No Trading & 0.49\% & 0.00\% & 0.00\% & 0.00\% & 0.00\% \\
Error & 0.00\% & 0.00\% & 0.00\% & 0.00\% & 0.00\% \\
\midrule
Optimal Herding Opportunities & 48.15\% & 42.39\% & 35.51\% & 51.93\% & 66.14\% \\
\bottomrule
\end{tabular}

        }
    }
\end{table}
\clearpage

\begin{table}[h!]
    \caption{
        \textbf{Trading behavior in AI laboratory with different payoffs}
        \vspace{10pt}\\
        \footnotesize{The table shows the distribution of decisions in the AI laboratory when varying the description of payoff in the prompt. First, payoffs are assumed to be worthless by setting a zero exchange rate between GBP and lire. Next, the stakes are increased by imposing a one million GBP per lire exchange rate. Finally, GBP is replaced by USD. Decisions are averaged across all sessions, trading periods, and LLMs in (a) Treatment I (without event uncertainty), (b) Treatment II (with event uncertainty), and (c) Treatment III (without price updating). \quotation{Rational} behavior represents cases where the informed trader chooses to buy upon receiving a white signal and sell upon receiving a blue signal. \quotation{Partial Rational} behavior represents cases where the  informed trader chooses to buy (sell) upon receiving a white (blue) signal and not trade upon receiving the other signal. \quotation{Cascade Trading} represents cases where the informed trader chooses the same trading action (buy or sell) regardless of the private signal. These decisions are decomposed into \quotation{Optimal Herding}, \quotation{Suboptimal Herding}, \quotation{Contrarian} behavior, and cases where the trade imbalance is zero (\quotation{Undetermined}). \quotation{Cascade No Trading} represents cases where the informed trader chooses not to trade regardless of the private signal. \quotation{Error} represents cases where the informed trader chooses to buy upon receiving a blue signal and sell upon receiving a white signal. The table also reports the frequency of trading periods where herding is optimal. }
    }
    \label{tab:table1_payoffs}
    \centering
    \footnotesize{
        \subfloat[Treatment I]{
        \begin{tabular}{lccc}
\toprule
 & 0 GBP per lire & 1M GBP per lire & 3 lire per USD \\
\midrule
Rational & 50.59\% & 52.33\% & 48.87\% \\
Partial Rational & 41.13\% & 35.46\% & 39.93\% \\
Cascade Trading & 7.88\% & 12.21\% & 8.07\% \\
\hskip15pt Optimal Herding & 0.00\% & 0.00\% & 0.00\% \\
\hskip15pt Suboptimal Herding & 0.00\% & 0.00\% & 0.00\% \\
\hskip15pt Contrarian & 7.88\% & 12.21\% & 8.07\% \\
\hskip15pt Undetermined & 0.00\% & 0.00\% & 0.00\% \\
Cascade No Trading & 0.39\% & 0.00\% & 3.12\% \\
Error & 0.00\% & 0.00\% & 0.00\% \\
\midrule
Optimal Herding Opportunities & 0.00\% & 0.00\% & 0.00\% \\
\bottomrule
\end{tabular}
}
        \\
        \emph{(Table continues on next page)}
    }
\end{table}
\clearpage

\begin{table}[h!]\ContinuedFloat
    \caption{
        \textbf{Trading behavior in AI laboratory with different payoffs} \emph{(continued)}
        \vspace{10pt}\\
    }
    \centering
    \footnotesize{
        \subfloat[Treatment II]{
        \begin{tabular}{lccc}
\toprule
 & 0 GBP per lire & 1M GBP per lire & 3 lire per USD \\
\midrule
Rational & 97.27\% & 95.21\% & 97.07\% \\
Partial Rational & 2.73\% & 3.91\% & 2.93\% \\
Cascade Trading & 0.00\% & 0.88\% & 0.00\% \\
\hskip15pt Optimal Herding & 0.00\% & 0.39\% & 0.00\% \\
\hskip15pt Suboptimal Herding & 0.00\% & 0.00\% & 0.00\% \\
\hskip15pt Contrarian & 0.00\% & 0.00\% & 0.00\% \\
\hskip15pt Undetermined & 0.00\% & 0.49\% & 0.00\% \\
Cascade No Trading & 0.00\% & 0.00\% & 0.00\% \\
Error & 0.00\% & 0.00\% & 0.00\% \\
\midrule
Optimal Herding Opportunities & 39.04\% & 34.49\% & 43.50\% \\
\bottomrule
\end{tabular}
}\\
        
        \subfloat[Treatment III]{
        \begin{tabular}{lccc}
\toprule
 & 0 GBP per lire & 1M GBP per lire & 3 lire per USD \\
\midrule
Rational & 99.90\% & 99.65\% & 99.21\% \\
Partial Rational & 0.10\% & 0.35\% & 0.16\% \\
Cascade Trading & 0.00\% & 0.00\% & 0.63\% \\
\hskip15pt Optimal Herding & 0.00\% & 0.00\% & 0.63\% \\
\hskip15pt Suboptimal Herding & 0.00\% & 0.00\% & 0.00\% \\
\hskip15pt Contrarian & 0.00\% & 0.00\% & 0.00\% \\
\hskip15pt Undetermined & 0.00\% & 0.00\% & 0.00\% \\
Cascade No Trading & 0.00\% & 0.00\% & 0.00\% \\
Error & 0.00\% & 0.00\% & 0.00\% \\
\midrule
Optimal Herding Opportunities & 35.08\% & 37.67\% & 62.95\% \\
\bottomrule
\end{tabular}

        }
    }
\end{table}
\clearpage

\begin{table}[h!]
    \caption{
        \textbf{Trading behavior in AI laboratory with prolonged experiments}
        \vspace{10pt}\\
        \footnotesize{The table shows the distribution of decisions in the AI laboratory when varying the length of the experiment (number of trading periods and number of independent sessions). Decisions are averaged across all sessions, trading periods, and LLMs in (a) Treatment I (without event uncertainty), (b) Treatment II (with event uncertainty), and (c) Treatment III (without price updating). \quotation{Rational} behavior represents cases where the informed trader chooses to buy upon receiving a white signal and sell upon receiving a blue signal. \quotation{Partial Rational} behavior represents cases where the  informed trader chooses to buy (sell) upon receiving a white (blue) signal and not trade upon receiving the other signal. \quotation{Cascade Trading} represents cases where the informed trader chooses the same trading action (buy or sell) regardless of the private signal. These decisions are decomposed into \quotation{Optimal Herding}, \quotation{Suboptimal Herding}, \quotation{Contrarian} behavior, and cases where the trade imbalance is zero (\quotation{Undetermined}). \quotation{Cascade No Trading} represents cases where the informed trader chooses not to trade regardless of the private signal. \quotation{Error} represents cases where the informed trader chooses to buy upon receiving a blue signal and sell upon receiving a white signal. The table also reports the frequency of trading periods where herding is optimal.}
    }
    \label{tab:table1_length}
    \centering
    \footnotesize{
        \subfloat[Treatment I]{
        \begin{tabular}{lccc}
\toprule
 & Baseline (4 sessions of 8 rounds) & 10 sessions of 8 rounds & 4 sessions of 20 rounds \\
\midrule
Rational & 61.00\% & 45.08\% & 52.69\% \\
Partial Rational & 29.48\% & 32.28\% & 33.23\% \\
Cascade Trading & 9.42\% & 20.26\% & 11.53\% \\
\hskip15pt Optimal Herding & 0.00\% & 0.00\% & 0.00\% \\
\hskip15pt Suboptimal Herding & 0.00\% & 0.00\% & 0.00\% \\
\hskip15pt Contrarian & 9.42\% & 20.26\% & 11.53\% \\
\hskip15pt Undetermined & 0.00\% & 0.00\% & 0.00\% \\
Cascade No Trading & 0.10\% & 2.38\% & 2.54\% \\
Error & 0.00\% & 0.00\% & 0.00\% \\
\midrule
Optimal Herding Opportunities & 0.00\% & 0.00\% & 0.00\% \\
\bottomrule
\end{tabular}
}
        \\
        \emph{(Table continues on next page)}
    }
\end{table}
\clearpage

\begin{table}[h!]\ContinuedFloat
    \caption{
        \textbf{Trading behavior in AI laboratory with prolonged experiments} \emph{(continued)}
        \vspace{10pt}\\
    }
    \centering
    \footnotesize{
        \subfloat[Treatment II]{
        \begin{tabular}{lC{3cm}C{3cm}C{3cm}}
\toprule
 & Baseline (4 sessions of 8 rounds) & 10 sessions of 8 rounds & 4 sessions of 20 rounds \\
\midrule
Rational & 97.36\% & 89.43\% & 94.45\% \\
Partial Rational & 2.64\% & 6.48\% & 5.55\% \\
Cascade Trading & 0.00\% & 4.04\% & 0.00\% \\
\hskip15pt Optimal Herding & 0.00\% & 0.33\% & 0.00\% \\
\hskip15pt Suboptimal Herding & 0.00\% & 0.00\% & 0.00\% \\
\hskip15pt Contrarian & 0.00\% & 3.67\% & 0.00\% \\
\hskip15pt Undetermined & 0.00\% & 0.04\% & 0.00\% \\
Cascade No Trading & 0.00\% & 0.03\% & 0.00\% \\
Error & 0.00\% & 0.02\% & 0.00\% \\
\midrule
Optimal Herding Opportunities & 36.56\% & 65.73\% & 37.19\% \\
\bottomrule
\end{tabular}
}\\
        
        \subfloat[Treatment III]{
        \begin{tabular}{lC{3cm}C{3cm}C{3cm}}
\toprule
 & Baseline (4 sessions of 8 rounds) & 10 sessions of 8 rounds & 4 sessions of 20 rounds \\
\midrule
Rational & 99.65\% & 99.82\% & 99.81\% \\
Partial Rational & 0.16\% & 0.03\% & 0.19\% \\
Cascade Trading & 0.00\% & 0.15\% & 0.00\% \\
\hskip15pt Optimal Herding & 0.00\% & 0.15\% & 0.00\% \\
\hskip15pt Suboptimal Herding & 0.00\% & 0.00\% & 0.00\% \\
\hskip15pt Contrarian & 0.00\% & 0.00\% & 0.00\% \\
\hskip15pt Undetermined & 0.00\% & 0.00\% & 0.00\% \\
Cascade No Trading & 0.20\% & 0.00\% & 0.00\% \\
Error & 0.00\% & 0.00\% & 0.00\% \\
\midrule
Optimal Herding Opportunities & 38.36\% & 73.38\% & 44.86\% \\
\bottomrule
\end{tabular}

        }
    }
\end{table}
\clearpage

\end{appendices}

\end{document}